\documentclass%
[aps,preprintnumbers,showpacs,showkeys,nofootinbib,superscriptaddress]{revtex4}
\usepackage{epsfig,graphicx,psfrag}
\usepackage{amssymb,amsmath,amsfonts,amsthm}
\usepackage{multirow}
\usepackage{color}
\usepackage{hyperref}

\graphicspath{{./Figs/}}                 
\newcommand{\psibar}{\ensuremath{\bar\psi}}
\newcommand{\chibar}{\ensuremath{\bar\chi}}

\newcommand{\Nt}{\ensuremath{N_\tau}}
\newcommand{\Ns}{\ensuremath{N_\sigma}}
\newcommand{\vev}[1]{\ensuremath{\left<#1\right>}}
\newcommand{\vevsub}[1]{\ensuremath{\left<#1\right>_{\mathrm{sub}}}}
\newcommand{\rnulla}{\ensuremath{\left ( \frac{r_0}{a} \right ) }}
\newcommand{\rchi}{\ensuremath{r_\chi / a }}
\newcommand{\Real}{\ensuremath{\mathrm{Re}}}

\newcommand{\Trace}{\ensuremath{\mathrm{Tr}}}

\newcommand{\chidof}{\ensuremath{\chi^2/\mathrm{dof}}}

\newcommand{\e}{\ensuremath{\mathrm{e}}}

\newcommand{\Eq}[1]{Eq.\,(\ref{#1})}

\newcommand{\Fig}[1]{Fig.\,\ref{#1}}

\newcommand{\Cite}[1]{Ref.\ \cite{#1}}
\newcommand{\Cites}[1]{Refs.\ \cite{#1}}

\newcommand{\Tab}[1]{Table \ref{#1}}

\newcommand{\Sec}[1]{Section \ref{#1}}

\newcommand{\cf}{cf.\ }
\newcommand{\ie}{i.~e.\ }
\newcommand{\eg}{e.~g.\ }
\newcommand{\wrt}{w.~r.~t.\ }

\newcommand{\links}{\textbf{Left}: }
\newcommand{\mitte}{\textbf{Middle}: }
\newcommand{\rechts}{\textbf{Right}: }

\newcommand{\ronetwofive}{\ensuremath{\mathcal{R}_5^{1/2}}}
\newcommand{\tauthree}{\ensuremath{\tau^3}}
\newcommand{\gammafive}{\ensuremath{\gamma_5}}

\def\lsi{\raise0.3ex\hbox{$<$\kern-0.75em\raise-1.1ex\hbox{$\sim$}}}
\def\gsi{\raise0.3ex\hbox{$>$\kern-0.75em\raise-1.1ex\hbox{$\sim$}}}

\newlength{\graphicswidth}
\begin{document}
\preprint{HU-EP-14/64,~SFB/CPP-14-103}

\title{The equation of state of quark-gluon matter from lattice QCD \\
       with two flavors of twisted mass Wilson fermions}

\author{Florian Burger}
\affiliation{Humboldt-Universit\"at zu Berlin, Institut f\"ur Physik, 
 12489 Berlin, Germany}

\author{Ernst-Michael Ilgenfritz}
\affiliation{Joint Institute for Nuclear Research, VBLHEP and BLTP, 
141980 Dubna, Russia} 

\author{Maria Paola Lombardo}
\affiliation{Istituto Nazionale di Fisica Nucleare, Sezione di Pisa, 
Largo Pontecorvo 3, I-56100 Pisa, and 
Laboratori Nazionali di Frascati, INFN, 
100044 Frascati, Roma, Italy}

\author{Michael M{\"u}ller-Preussker} 
\affiliation{Humboldt-Universit\"at zu Berlin, Institut f\"ur Physik, 
12489 Berlin, Germany}

\collaboration{tmfT collaboration}
\noaffiliation

\date{December 21, 2014}


\begin{abstract}
We report on lattice QCD results for the thermodynamic equation of state of 
quark-gluon matter obtained with $N_f=2$ degenerate quark flavors. 
For the fermion field discretization we are using the Wilson twisted mass 
prescription. Simulations have been carried out at three values of the bare 
quark masses corresponding to pion masses of $\sim 360, \sim 430$ and 
$\sim 640$ MeV. We highlight the importance of a good control of the lattice 
cutoff dependence of the trace anomaly which we have studied at several
values of the inverse temperature $T^{-1}=a N_\tau$ with a time-like lattice 
extent up to $N_\tau=12$. We contrast our results with those of other groups 
obtained for $N_f=0$ and $N_f=2+1$. At low temperature we also confront them 
with hadron resonance gas model predictions for the trace anomaly. 
\end{abstract}

\maketitle

\section{Introduction} \label{sec:intro}

Lattice QCD investigations of the (pseudo-) critical behavior of 
quark and gluon matter at varying  temperature have been carried out over many
years by several groups employing various improved discretization prescriptions.
The special and very demanding task to determine the thermodynamic equation 
of state (EoS) has reached the physical point, i.e. realistic up-, 
down- and strange-quark masses. For this aim highly improved staggered fermion 
discretizations have been employed as reported by the Budapest-Wuppertal group 
\cite{Borsanyi:2010cj,Borsanyi:2013bia} and the HotQCD collaboration 
\cite{Bazavov:2014pvz} (both for $N_f=2+1$ dynamical quark degrees of freedom).  
The staggered fermion approach is most effective from the computational point 
of view. However, one pays the price of a theoretical uncertainty by 
applying the rooting trick for the fermionic determinant in order to reduce
unwanted (taste) degrees of freedom. 

The (improved) Wilson fermion approach is theoretically safe but 
computationally very demanding and has arrived at the physical point for 
the zero-temperature case with a corresponding delay \cite{Abdel-Rehim:2013yaa}. 
For thermodynamic applications this limit probably will still need some more 
time. Thermodynamics with improved Wilson quarks has been studied for two quark 
flavors more than one decade ago by the CP-PACS collaboration 
\cite{AliKhan:2001ek}. At that time only a very small lattice extent in the 
Euclidean time direction was feasible ($N_{\tau}=4,~\mathrm{or}~6$). Therefore, 
the results were strongly influenced by lattice artifacts. The DIK collaboration
continued this effort by enlarging $N_\tau$ up to 14 lattice units 
\cite{Bornyakov:2009qh,Bornyakov:2011yb}. More recently improved Wilson 
fermions were studied with $N_f = 2+1$ on large lattices by the WHOT 
collaboration \cite{Umeda:2012er} and the Budapest-Wuppertal group 
\cite{Borsanyi:2011kg,Borsanyi:2012uq}.
Let us also mention attempts to study lattice QCD at nonzero temperature
with chirally perfect fermion approaches like the domain wall ansatz 
\cite{Cheng:2008ge,Cheng:2009be,Chiu:2013wwa} and overlap fermions
\cite{Bornyakov:2013eya}.

Simulations with a dynamical charm quark, which is expected to be relevant 
above temperatures of 400 MeV are also in the course of being performed by 
the Budapest-Wuppertal \cite{Ratti:2013uta} and MILC \cite{Bazavov:2013pra} 
collaborations employing staggered discretizations and by us with 
Wilson twisted mass quarks in a fixed scale study \cite{Burger:2013hia}.
Recent reviews of the whole subject can be found in 
\cite{DeTar:2009ef,DeTar:2011nm,Levkova:2012jd,Philipsen:2012nu,
Petreczky:2012rq,Lombardo:2012ix}.

In addition to the investigations mentioned above, the analysis of the
two-flavor model has several reasons of interest. The crossover region and 
the issue of universality in the chiral limit have been investigated by us with 
twisted mass Wilson fermions \cite{Burger:2011zc,Burger:2012zz}, with clover 
improved fermions in \cite{Brandt:2013mba} and with $N_f=2$ staggered flavors 
with imaginary chemical potential \cite{Bonati:2014kpa}.

Here, we present our results for the EoS in the two-flavor case with 
twisted mass fermions. Preliminary results can be found in 
\cite{Burger:2012zz}. From a technical viewpoint, we want to see how
twisted mass Wilson fermions perform in the determination of the EoS.
Given satisfactory performance we may aim at studying the flavor dependence of 
the EoS in the critical region by comparing with the quenched and the 
$N_f=2+1$ cases.

Our main observable is the trace anomaly (also called interaction measure)
\begin{equation}
  I = \epsilon - 3 p  
    = T^5 \frac{\partial}{\partial T} \left (  \frac{p}{T^4}  \right ) 
  \label{eq:pressurederiv}
\end{equation}
related to the partition function by a total derivative with respect to 
the lattice spacing $a$
\begin{equation}
  I = - \frac{T}{V} \frac{d \ln Z}{d \ln a}.  
  \label{eq:traceanomalyevaluation}
\end{equation}
For the calculation of the temperature dependence of the pressure $p(T)$ and 
the energy density $\epsilon(T)$ we will employ the integral method 
(see \eg \cite{DeTar:2009ef})
according to which the pressure can be evaluated by integrating 
\Eq{eq:pressurederiv}
\begin{equation}
  \frac{p}{T^4} - \frac{p_0}{T_0^4} = 
  \left.\int_{T_0}^{T} d\tau \frac{\epsilon - 3 p}{\tau^5} \right |_\mathrm{LCP}
\label{eq:pressureintegral}
\end{equation}
along a line of constant physics (LCP). The lower bound of the integration 
has to be set at a sufficiently low temperature $T_0$ in such a way that 
$p_0$ is close to zero and can be neglected. Alternatively it may be
set from a hadron resonance gas (HRG) model analysis.

In \Sec{sec:setup} we describe our lattice setup using twisted mass fermions
and the tree-level Symanzik improved gauge action followed by the outline
of the scale setting prescription in \Sec{sec:scalesetting}. In \Sec{sec:lcp}
we are describing the lines of constants of physics along which the temperature
integration will be carried out. Since we have redone our scale setting in 
comparison to our previous work \Cite{Burger:2011zc}, and since we have added a
new (higher) pion mass value we present a new computation of the pseudo-critical 
temperature in \Sec{sec:transtemp}. \Sec{sec:betafun} provides all details for
the computation of the $\beta$-function and prefactors of the 
fermionic contributions to the trace anomaly. Since we have to subtract 
$T=0$ results, which are not available for all parameter values discussed,
we present the outcome of corresponding interpolations in \Sec{sec:t0int}.  
Finally, our results for the trace anomaly, pressure and energy density 
as functions of the temperature are shown in Sections \ref{sec:traceanomaly} 
and \ref{sec:thermo}. 
In Section \ref{sec:hrg} we show how to use an appropriate modification
of the hadron resonance gas model to estimate the low temperature contribution
to the pressure. Finally, in Section \ref{sec:comparison} 
we contrast our $N_f=2$ determination with the quenched and 
$N_f=2+1$ results. In \Sec{sec:conclusions} we will draw the conclusions.
In Appendix \ref{sec_m0symanzik} we collect the arguments, why one
of the contributions to the trace anomaly vanishes in the continuum limit
and therefore can be neglected from the beginning. Appendix \ref{sec:simtables}
lists all tables of simulation parameters and of unrenormalized data
for the Polyakov loop and the chiral condensate.


\section{Lattice Twisted Mass Setup} \label{sec:setup}
For the present study of $N_f = 2$ thermodynamics we have been relying on 
the twisted mass lattice quark action for two flavors of mass-degenerate quarks
\begin{equation}
S^{\mathrm{tm}}_f[U,\psi,\psibar] 
= \sum_{x,y} \chibar(x) \Big{(} a m \delta_{x,y} + D_\mathrm{W}(x,y)[U] 
+  i a \mu \gamma_5\tau^3 \delta_{x,y} \Big{)} \chi(y) \;. 
\label{eq:tmaction}
\end{equation}
The Wilson discretization of the covariant derivative is given by
\begin{equation}
 D_\mathrm{W}(x,y)[U] =  4 \delta_{x,y} + \frac{1}{2} 
        \sum_\mu \left ( 1 - \gamma_\mu  \right ) U_\mu(x)\delta_{y, x+a\hat \mu}
   + \left ( 1 + \gamma_\mu \right ) U^{\dagger}_\mu(x-a\hat \mu)  
     \delta_{y, x-a\hat \mu} \;,
\label{eq:covderi_wilson}
\end{equation}
where the usual Wilson parameter has been put $r \equiv 1$.
Via $\kappa \equiv \left ( 2 a m + 8 \right )^{-1}$ the bare (untwisted) 
quark mass $m$ is related to the hopping parameter $\kappa$ which 
has been set to its coupling dependent critical value $\kappa_c(\beta)$ as 
determined by the European Twisted Mass Collaboration (ETMC) 
\cite{Boucaud:2008xu} and suitably interpolated to the coupling values 
used in this study \cite{Burger:2011zc}. 

The gauge action is discretized with a tree-level Symanzik improved action
\begin{equation}
S_g^{\mathrm{tlSym}}[U] = \beta  \Big{(} c_0 \sum_{P} \lbrack 1 - 
\frac{1}{3} \Real \Trace \left ( U_{P} \right ) \rbrack  
 + c_1 \sum_{R} \lbrack  1 - 
\frac{1}{3} \Real \Trace  \left ( U_{R} \right ) \rbrack \Big{)} 
\label{eq:tlsym}
\end{equation}
with $c_0=5/3$ and $c_1=-1/12$ and sums extending over all plaquettes ($P$) and 
all planar rectangles ($R$) attached to each lattice site in positive 
directions, respectively. 

We have simulated three values of the pion mass $m_\pi \sim 360$, $\sim 430$ and 
$\sim 640$ MeV, refered to as the B-, C- and D-mass in what follows.  
In each case several values of $N_\tau$ ranging from $N_\tau=4$ to $N_\tau=12$ 
have been simulated, 
see Tables \ref{tab_Bensemble}, \ref{tab_Censemble}, and \ref{tab_Densemble}
in Appendix \ref{sec:simtables}.  
The data of the B- and the 
C-mass have already partly been used in \Cite{Burger:2011zc} for the study of 
the chiral limit of the transition
\footnote{
In \Cite{Burger:2011zc} an even smaller mass has been considered (called 
``A-mass''). We have not included it here, since for the EoS required $T=0$ 
simulations would run into metastable states of the bulk transition occuring 
at sufficiently small $\beta$. A way out would be to keep at larger 
$\beta$-values, i.e. to describe the crossover region with a time-like 
lattice extent $N_\tau \ge 14$. Such large lattice sizes go
beyond the scope of the present investigation.}. 
Preliminary results for the EoS in our setup have been reported in 
\Cite{Burger:2012zz}.

\section{Scale Setting} \label{sec:scalesetting}

The lattice scale is set using the physical value of the Sommer scale $r_0$ 
determined by ETMC from the nucleon mass in \Cite{Alexandrou:2010hf}. For the 
ETMC generated gauge ensembles the values for $r_0/a$ have been published 
in \Cite{Baron:2009wt}. Table \ref{tab_T0_ETM} in Appendix \ref{sec:simtables} 
lists the ETMC ensembles we have used and analysed in this work.
On additionally generated $T=0$ gauge ensembles, see Table \ref{tab_T0}, 
$r_0/a$ has been determined from the Euclidean time dependence of the 
static potential $V(x_4)$, which was extracted from  
timelike Wilson loops as described in \Cite{Baron:2009wt}. 
The latter have been evaluated on HYP- and APE-smeared configurations. 
In order to reduce lattice artefacts in the determination of $V(x_4)$ 
at low values of $\beta$
we employ a tree-level improved definition for the spatial separation 
as in \Cite{Jansen:2011vv} for $\beta \le 3.76$. We perform the chiral limit 
extrapolation of $r_0/a$ for obtaining $\rchi$ both with assuming either a 
linear dependence on the bare quark mass $\mu$ or a pure quadratic dependence 
without a linear piece. Half of the resulting difference of $r_\chi$ is 
taken as a systematic error and is added to the statistical error of $\rchi$.
The quadratically extrapolated values have entered our fits as central values.
For setting the scale and also for evaluating the $\beta$-function
(\cf Section \ref{sec:betafun}) we proceed by fitting $\rchi$ and the data 
shifted by $\pm 1 \sigma$ with the following ansatz:
\begin{equation}
  (\rchi) (\beta) = 
  \frac{1 + n_0 R(\beta)^2}{d_0 \, a_{\mathrm{2loop}}(\beta) \left (1  + 
                         d_1 R(\beta)^2 \right ) } \;.
 \label{eq:betafunfit}
\end{equation}
The ratio $R(\beta) \equiv a_{\mathrm{2loop}}(\beta)/
a_{\mathrm{2loop}}(\beta_{\rm{ref}})$ is defined in terms of the 
two-loop renormalization group formula for the (dimensionless) scale 
$a_{\mathrm{2loop}}(\beta)$ at vanishing fermion mass
\begin{equation}
      a_{\mathrm{2loop}}(\beta)  = 
               \left ( \frac{6 \beta_0}{\beta} \right )^{-\beta_1/2 \beta_0^2} 
                                 \exp \left ( -\frac{\beta}{12 \beta_0} \right)  
  \label{eq:twoloop}
\end{equation}
with the first two (universal) coefficients of the perturbative $\beta$-function 
$\beta_0=(11-2 N_f/3)/(4 \pi)^2$ and $\beta_1=(102 - \frac{38}{3} N_f)/(4 \pi)^4$. 
In our analyse we have chosen $\beta_{\rm{ref}} = 3.9$ for an intermediate 
reference scale 
and have checked that the results do not depend on this choice. We add the 
maximal deviation from the fits to the data of upper and lower error bands 
from the central fit to the statistical error in quadrature.
In order to account for the systematic error associated with the specific 
choice of a fit function we have performed the fits 
either setting $d_1 \equiv 0$ or keeping it 
as a free parameter in the fit, the former 
representing our central fit. We propagate the resulting difference of the 
such fitted $\beta$ dependences of $\rchi$ to all subsequent 
analyses such as the $\beta$-function and 
other scale dependent quantities.

\begin{figure}[tb] 
\hfill
\includegraphics[width=0.45\textwidth]{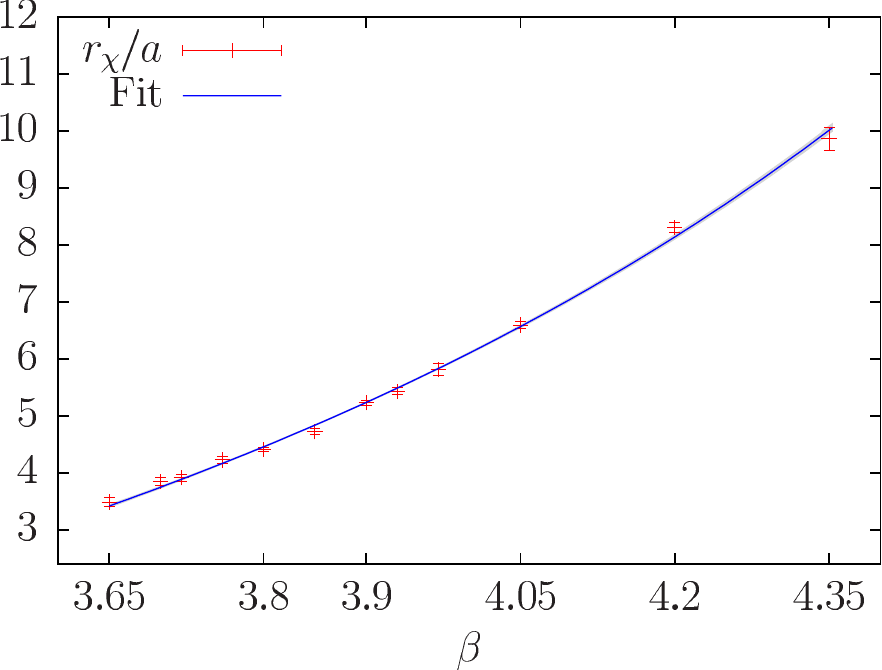} \hfill
\includegraphics[width=0.45\textwidth]{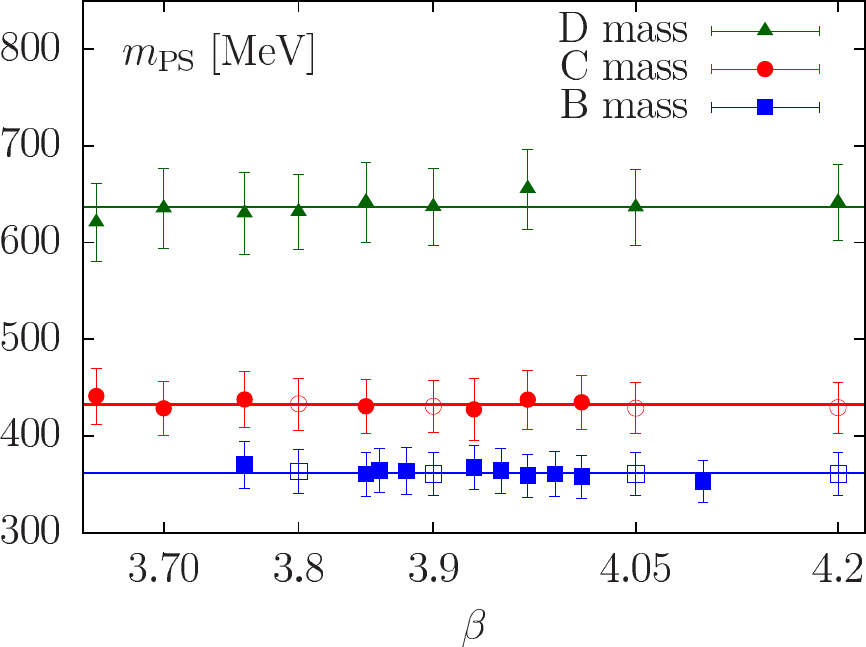} 
\hfill
\caption[]{\links Chirally extrapolated Sommer scale $\rchi$ and 
a fit using \Eq{eq:betafunfit}. \rechts Charged pion mass in physical 
units for the three ensembles together with a constant fit over 
all data points. Open symbols denote data that has been interpolated
using ETMC results.}
\label{fig_r0fit}
\end{figure}

Allowing for three fit parameters at maximum we use a lower number of 
parameters compared to \Cite{Cheng:2007jq}, 
since the functional dependence of $\rchi$ on $\beta$ is rather mild, and 
we have less data points to fit. In \Fig{fig_r0fit} we present the fit which -
with twelve data points for $\rchi$ at our disposal -
yields a good $\chi^2/\mathrm{dof} = 1.6$. The fitted parameters read as 
follows:
\begin{center}
\begin{tabular}{c|c|c|c|c}
  Fit 	& $n_0$ 	&  $d_0$ 	& $d_1$ & $\chi^2/\mathrm{dof} $\\
 \hline
 1  	& $-0.1096(75)$ 	&  $13.04(12)$  & \textbf{0}	& 1.6\\
 2  	& $0.35(30)$ 		&  $12.27(44)$  & $0.62(43)$	& 1.1\\ 
\end{tabular}
\end{center}
The temperature in physical units is then estimated from the fit using 
\begin{equation}
    T (\beta) \ \ [\mathrm{MeV}] = 
  \frac{\left (r_\chi/a \right )(\beta)}{N_\tau \ \ r_0} 
\end{equation}
and taking 
$r_0 =  0.462(28)$ fm from \Cite{Alexandrou:2010hf}  
as input. The uncertainty of the temperature evaluated in this manner 
is of the order of 4 \% throughout the whole temperature range.

\section{Lines of Constant Physics} \label{sec:lcp}

The calculation of the pressure by means of integrating
\Eq{eq:pressureintegral} has to be done on the LCP. To this end we have 
fixed the mass of the (charged) pion $m_{\mathrm{PS}}$ 
to three constant values by tuning the bare quark 
mass by means of the $\beta$-function. The quality of mass tuning is 
shown in the right panel of \Fig{fig_r0fit} for the three masses together
with a constant fit over the whole range of the coupling. 
For the B-mass such a figure has already been shown in \Cite{Burger:2012zz} 
- note however that due to the updated value of $r_0$ the curve shown 
there is now slightly shifted. From the fit for the B-, C- and D-masses 
we obtain the values $m_{\mathrm{PS}} = 362(2)$ MeV, $433(2)$ MeV and 
$637(4)$ MeV, respectively. 
In the plot we do not show data points obtained at $\beta=4.35$. In this
case the box size is very small which leads to an over-estimation
of the pseudoscalar masses by about 20~\%.

\section{Thermal Transition Temperature} \label{sec:transtemp}

Since the scale setting has changed with respect to our previous study in 
\Cite{Burger:2011zc} and one more pion mass has been simulated,
by applying the same methods we have conducted a new determination of 
the pseudo-critical chiral temperature ($T_\chi$) and 
- what we conditionally call - the ``deconfinement'' temperature
($T_\mathrm{deconf}$).

\begin{figure}[htb] 
\centering
\hfill
\includegraphics[width=0.3\textwidth]{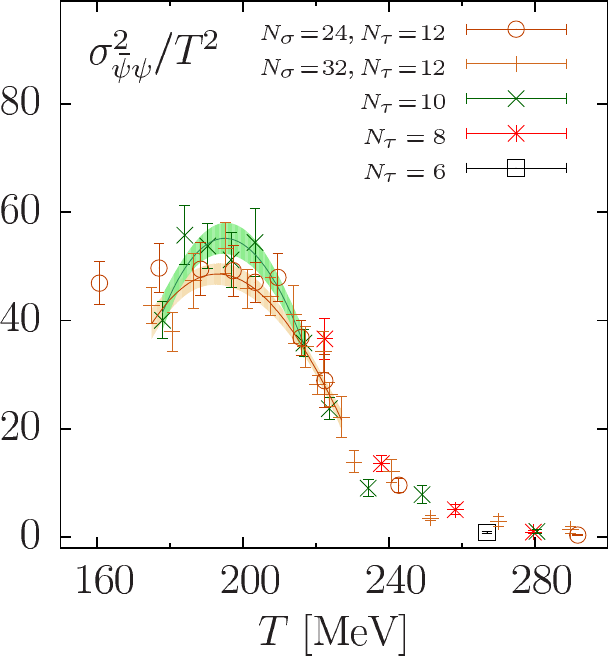}
\hfill
\includegraphics[width=0.3\textwidth]{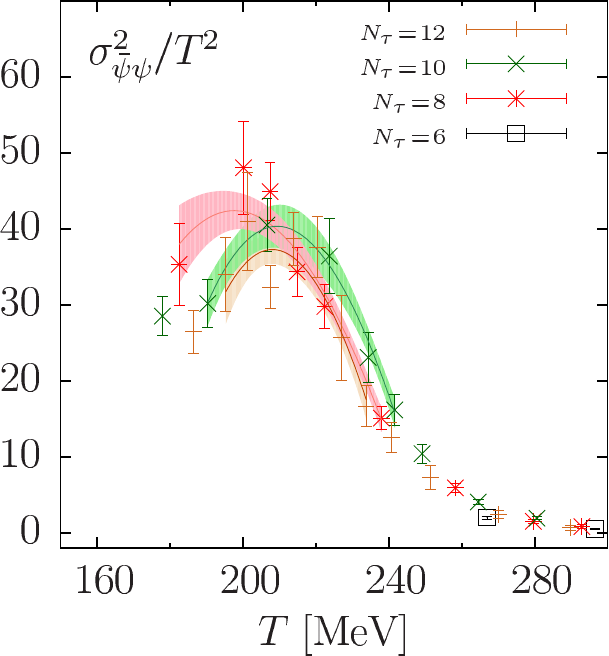}
\hfill
\includegraphics[width=0.3\textwidth]{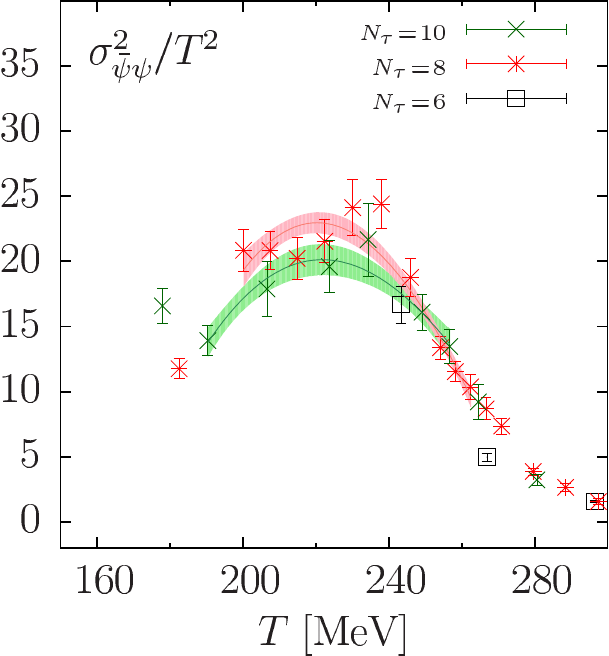}
\hfill
 \caption{The disconnected part of the chiral susceptibility in the 
crossover temperature range together with Gaussian fit curves 
versus temperature. \links For the B mass, \mitte C mass, and 
\rechts D mass. 
 }
\label{fig_pbpsusc}
\end{figure}

The chiral temperature $T_{\chi}$ is obtained from fitting Gaussian functions 
\begin{equation}
  G(\beta) = a_G + b_G \cdot \e^{-c_G (\beta - \beta_\chi)^2} 
\end{equation}
to the variance of the chiral condensate over configurations, i.e. to (the 
disconnected part of) the chiral susceptibility, which shows a maximum in the 
expected crossover temperature region. The fits are performed in the bare 
coupling, and the pseudo-critical coupling $\beta_\chi$ at the center of the 
Gaussian function is then converted into physical units. In \Fig{fig_pbpsusc} 
we show the chiral susceptibility normalised by the squared temperature 
for the three simulated pion masses as well as for several values of 
$N_\tau$. Note that the data has not been renormalised yet. 
In addition, in each case we show also the fit curves for the 
finest discretization of the Euclidean time extent (largest $N_\tau$ 
available). For the smallest mass we can check whether the 
thermodynamic limit can be considered to be satisfactorily achieved
as lattices of a smaller extent ($N_\sigma = 24$) are available
for comparison. Since - within errors - the susceptibility data for the 
smaller volume is compatible with the data obtained on the larger
volume ($N_\sigma = 32$), we conclude that $T_{\chi}$ is not 
affected by finite-size scaling effects as one should expect for a 
crossover phenomenon. We have restricted the Gaussian fits to data obtained 
in the larger volume.  

\begin{figure}[htb] 
\centering
\hfill
\includegraphics[width=0.3\textwidth]{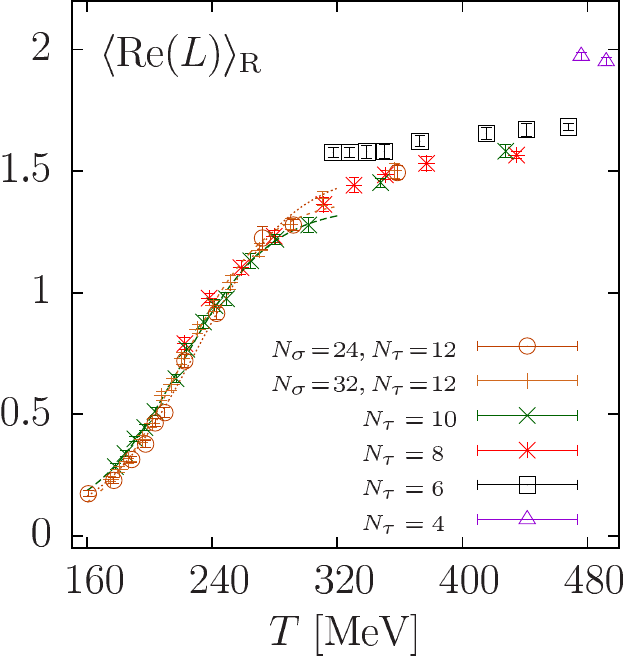}
\hfill
\includegraphics[width=0.3\textwidth]{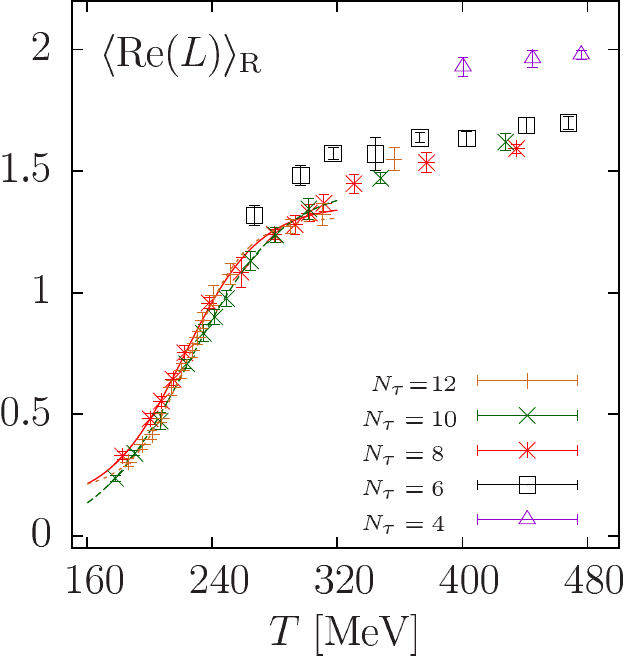}
\hfill
\includegraphics[width=0.3\textwidth]{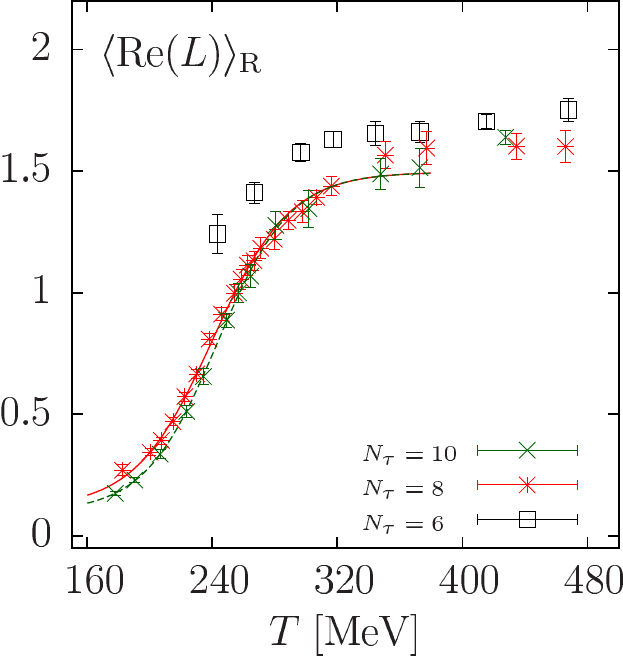}
\hfill
 \caption{The renormalised real part of the Polyakov loop versus temperature. 
  \links For the B mass, \mitte C mass, and \rechts D mass. Also shown are 
fits with $P(T)$ according to \Eq{eq:fit_P} (where evaluated).
 }
\label{fig_polyren}
\end{figure}

The deconfinement temperature $T_\mathrm{deconf}$ is estimated from the 
renormalised real part of the Polyakov loop $\vev{\Real (L)}_R$ obtained 
by multiplicative renormalisation using the static potential $V$ at zero 
temperature and distance $r_0$,
\begin{equation}
  \vev{\Real(L)}_R = \exp{(V(r_0)/2T)} \vev{\Real(L)}  
\equiv Z_L  \vev{\Real(L)} \,.
\label{eq:renormpoly}
\end{equation}
It is read off from the inflection point of a hyperbolic tangent function
\begin{equation}
  P(T) =  a_P + b_P \cdot \tanh{\left ( c_P (T-T_\mathrm{deconf}) \right )}
\label{eq:fit_P}
\end{equation}
by fitting the renormalized real part of the Polyakov loop 
while ignoring the uncertainty in the temperature scale.
For $\Nt \ge 8 $, $\vev{\Real(L)}_R$ shows only small lattice artefacts 
while they are sizable for $\Nt = 4$ and $\Nt=6$ as can be seen in 
\Fig{fig_polyren}. The fitted values for the deconfinement temperature 
are listed in the last column of \Tab{tab:tpc}. The first error indicates 
the statistical error while the second denotes the uncertainty of the 
temperature from the scale setting at the fitted $T_\mathrm{deconf}$.

In what follows, for the ``pseudo-critical temperature'' $T_c$ we will 
always use  $T_c \equiv T_\chi$ at the largest $N_\tau$ available. 

\begin{table}
\centering
\begin{tabular}{cccccc}
Ensemble & $m_\mathrm{PS}$\,[MeV]  & $\Nt \times \Ns^3$ & $\beta_c$ & 
                       $T_\chi$\,[MeV] &  $T_\mathrm{deconf}$\,[MeV]\\
\hline
\hline
 B   & $\sim 360$   & $12 \times 32^3$ & 3.92(1)    & 193(13)  & 219(3)(14)\\
     & 		    & $12 \times 24^3$ & -	    &    -     & 223(3)(14)\\
     & 		    & $10 \times 32^3$ & 3.82(1)    & 195(13)  & 219(4)(14) \\
\hline
 C   & $\sim 430$   & $12 \times 32^3$ & 3.97(2)    & 208(14)  & 225(3)(14)\\
     & 		    & $10 \times 32^3$ & 3.86(1)    & 209(14)  & 225(4)(14)\\ 
     & 		    & $8  \times 28^3$ & 3.69(3)    & 198(15)  & 219(6)(14)\\     
\hline     
 D   & $\sim 640$   & $10 \times 24^3$ & 3.90(3)    & 229(16)  & 244(3)(15)\\
     & 		    & $8 \times 20^3$  & 3.75(1)    & 225(15)  & 240(2)(15)\\      
\end{tabular}
\caption{List of extracted values for the pseudo-critical temperatures 
$T_\chi$ and $T_\mathrm{deconf}$.}
\label{tab:tpc}
\end{table}

\section{Trace anomaly and scale dependence of the partition function}
\label{sec:betafun}

The computation of the trace anomaly according to
\Eq{eq:traceanomalyevaluation} requires to evaluate the 
derivatives of the partition function with respect to the bare parameters  
$\kappa$, $a \mu$ and $\beta$. On the LCP the bare hopping parameter as well 
as the twisted mass are in turn functions of the gauge coupling.

Employing the following derivatives - which in analogy to the well-known 
$\beta$-function let us call $B$-functions - 
\begin{equation}
  B_\beta =  a \frac{d \beta}{d a} \;, 
  \quad \quad 
  B_\mu = \frac{1}{(a \mu)}\frac{\partial (a \mu)}{\partial \beta} \;, 
 \quad \quad 
  B_\kappa =  \frac{\partial{\kappa_c}}{\partial{\beta}}\;, 
 \quad \quad  
  B_m =   -\frac{1}{(a m)}\frac{1}{ (2 \kappa_c)^2} B_\kappa 
  \label{eq:deri}
\end{equation}
as well as the explicit form of our lattice 
action (\Eq{eq:tmaction} and \Eq{eq:tlsym}) we arrive at 
\begin{equation}
  \begin{split}
    \frac{I}{T^4} &= -N_\tau^4 B_\beta \frac{1}{N_\sigma^3 N_\tau} \Bigg \{ 
         \vevsub{\frac{c_0}{3} \sum_P \Real \Trace U_P}  +  
         \vevsub{\frac{c_1}{3} \sum_R \Real \Trace U_R} \\
      &\hphantom{-N_\tau^4 B_\beta \frac{1}{N_\sigma^3 N_\tau} \Bigg \{} 
      - (a \mu) B_\mu   \vevsub{\sum_x \bar \chi_x i \gamma_5 \tau^3 \chi_x}
      - (a m) B_{m}    \vevsub{\sum_x \bar \chi_x \chi_x}  \Bigg \} \;.
   \end{split}
\label{eq:traceanomaly}
\end{equation}
The expectation values are defined with an implicit subtraction of the 
corresponding expectation value at $T=0$ in order to render them finite in 
the ultraviolet: 
\begin{equation}
    \vevsub{\ldots} \equiv \vev{\ldots}_{T>0} - \vev{\ldots}_{T=0} \,.
\label{eq:vevsub}
\end{equation}
For later use (see e.g. Appendix B) we will abbreviate the gauge part 
in form of plaquette and rectangle contributions as 
\begin{equation}
S_g= \frac{1}{N_\tau N_\sigma^3} 
   \left ( c_0 \frac{1}{3} \sum_{P} \Real \Trace \left ( U_{P} \right ) 
  - c_1 \frac{1}{3} \sum_{R} \Real \Trace \left ( U_{R} \right ) \right )
\end{equation}
and the condensate contribution as
\begin{equation}
 S_f = -\frac{1}{N_\tau N_\sigma^3} 
        \sum_x \bar \chi_x i \gammafive \tauthree \chi_x\,.
\end{equation}
Since we are partly relying on the available $T=0$ ETMC data, for taking 
these subtractions it is necessary to interpolate the data in the mass 
as well as in the coupling. We shall discuss the strategy in detail in 
section \ref{sec:t0int}.
\begin{figure}[htb] 
\centering
\hfill
 \includegraphics[width=0.45\textwidth]{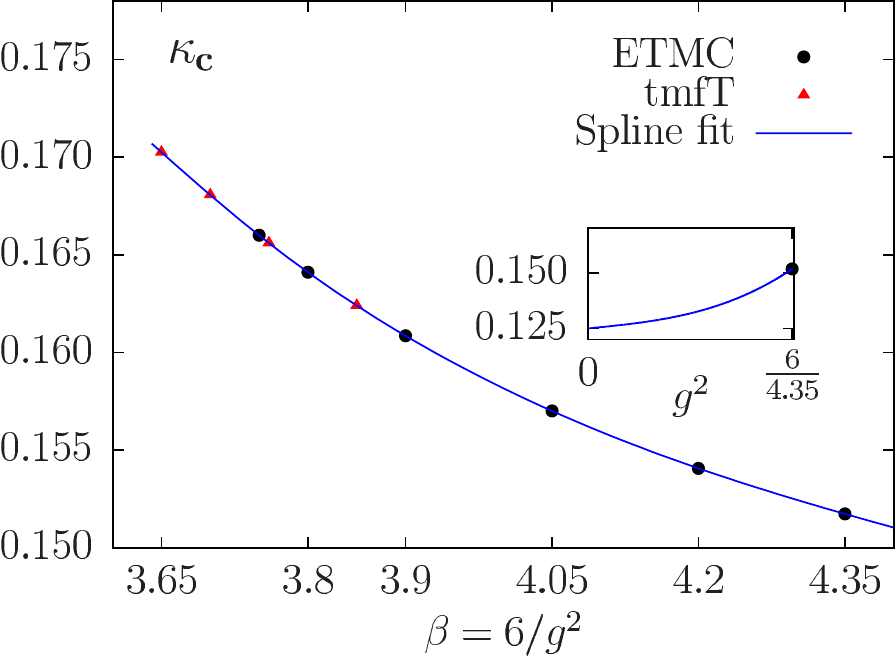} 
\hfill
 \includegraphics[width=0.45\textwidth]{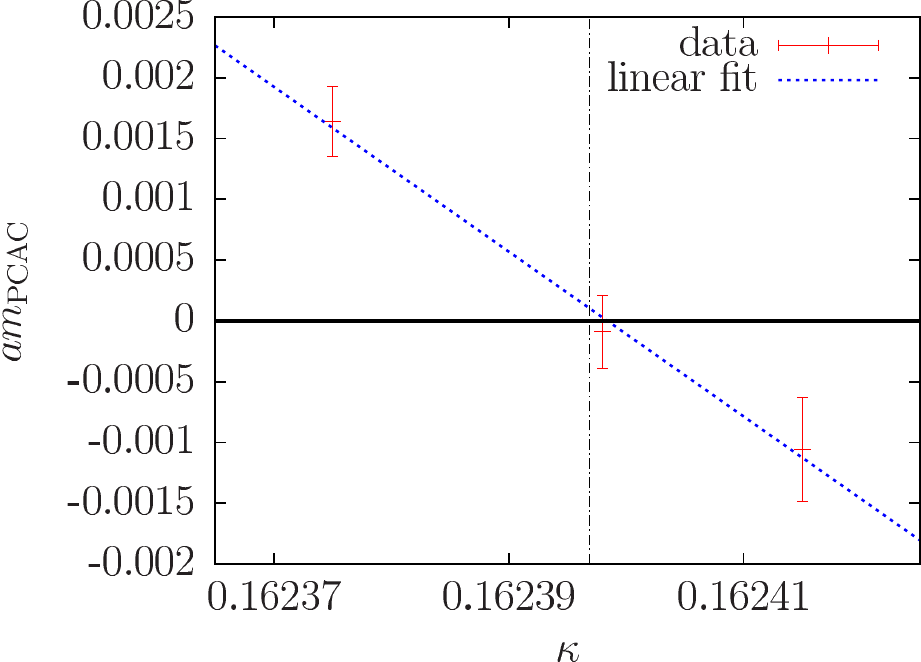} 
\hfill
 \caption{\links The dependence of the critical hopping parameter 
  $\kappa_c$ on the coupling $\beta$. The curve represents a Pad\'{e} 
  interpolation. In the inlaid figure we show the interpolation for 
  asymptotically large $\beta=6/g^2$, \ie small $g^2$, where the fit has 
  been constrained to $\kappa_c(g \equiv 0)=1/8$ representing the 
  asymptotic free limit value. \\
  \rechts The PCAC mass as a function of the hopping parameter 
  $\kappa$ around the critical value at $\beta = 3.85, a \mu = 0.006$.
  This corresponds to a check of the validity the spline interpolation 
  of $\kappa_c$ (see text for details).
  }
\label{fig_kcbeta}
\end{figure}

The untwisted quark mass related function $B_m$ is calculated using ETMC input. 
To this end we employ the coupling dependence of the critical hopping 
parameter $\kappa_c$, using the prescription indicated in \Eq{eq:deri}. 
We have fitted this dependence with a spline ansatz
which is shown in \Fig{fig_kcbeta}. In the low coupling region we have added 
several further estimates of $\kappa_c$. Since we have started the tuning 
at the largest mass, it was necessary to refine the tuning for smaller 
masses, such that for couplings below $\beta = 3.78$ slightly 
varying values for the critical hopping parameter have been simulated at a 
given coupling but at varying twisted mass, while ETMC kept $\kappa_c$ 
fixed in this case. For one value of the coupling ($\beta = 3.85, 
a \mu = 0.006$ ) we have conducted a check how well the interpolation 
works in determining $\kappa_c$ at an intermediate coupling. To this end 
we have simulated three values of $\kappa$ in the vicinity of the value 
predicted by the interpolation and have evaluated the PCAC mass which upon  
vanishing acts as a criterion for maximal twist \cite{Boucaud:2008xu}. 
The PCAC mass as well as a linear fit to the data are shown in the 
right panel of \Fig{fig_kcbeta}. The critical value of $\kappa$ as 
predicted from the interpolation is indicated by the vertical line. What 
can also be seen in the figure is that a slight mistuning leading to
a deviation of $\mathcal{O}(10^{-3})$ in the PCAC mass results in a 
value of $\kappa$ being off by $\mathcal{O}(10^{-5})$ from its critical 
value. We thus conclude that the error on the critical $\kappa$ 
dependence is very small and we thus neglect it in the further analysis,  
especially for evaluating $B_m$  which we take explicitly
from the derivative of $\kappa_c(\beta)$ using the fitted spline 
interpolation. For this we have used $\kappa_c$ as determined at the 
lowest mass at given coupling $\beta$.

\begin{figure}[htb] 
\centering
  \includegraphics[width=0.3\textwidth]{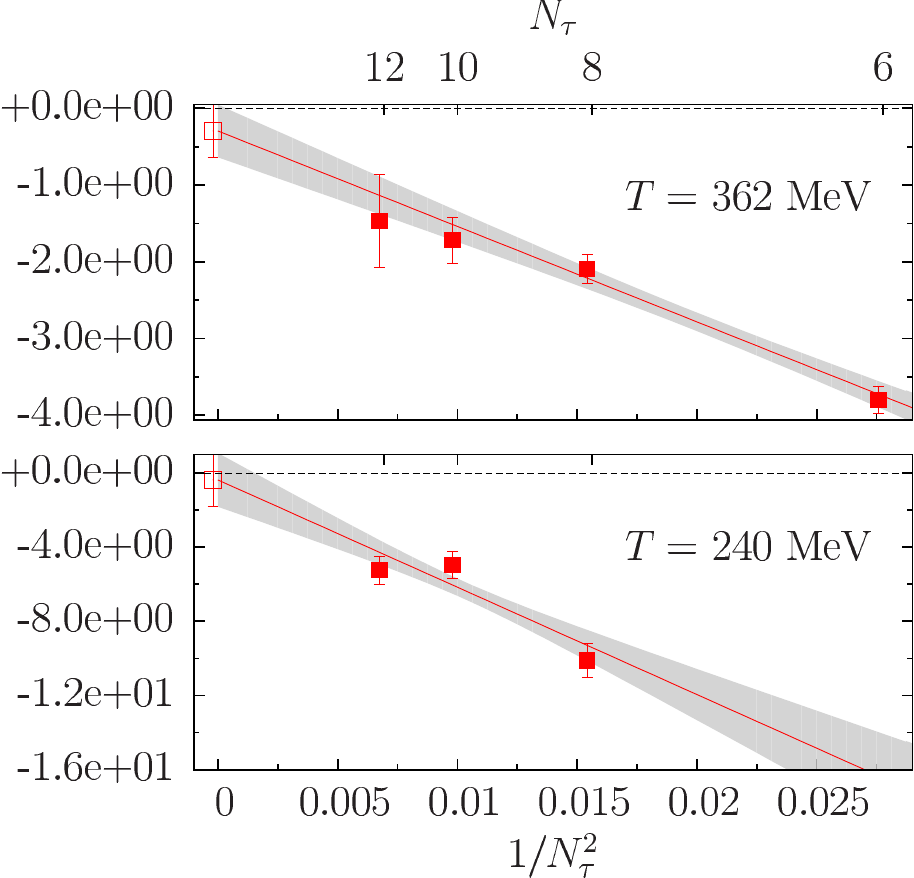}
  \includegraphics[width=0.3\textwidth]{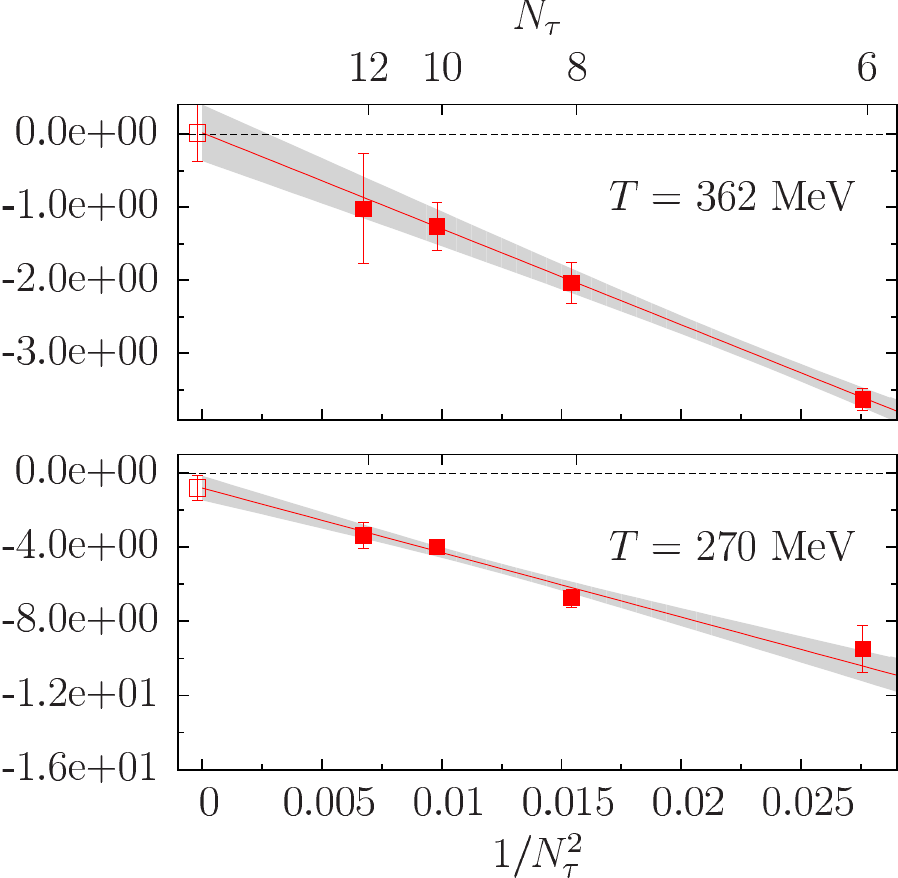}
  \includegraphics[width=0.3\textwidth]{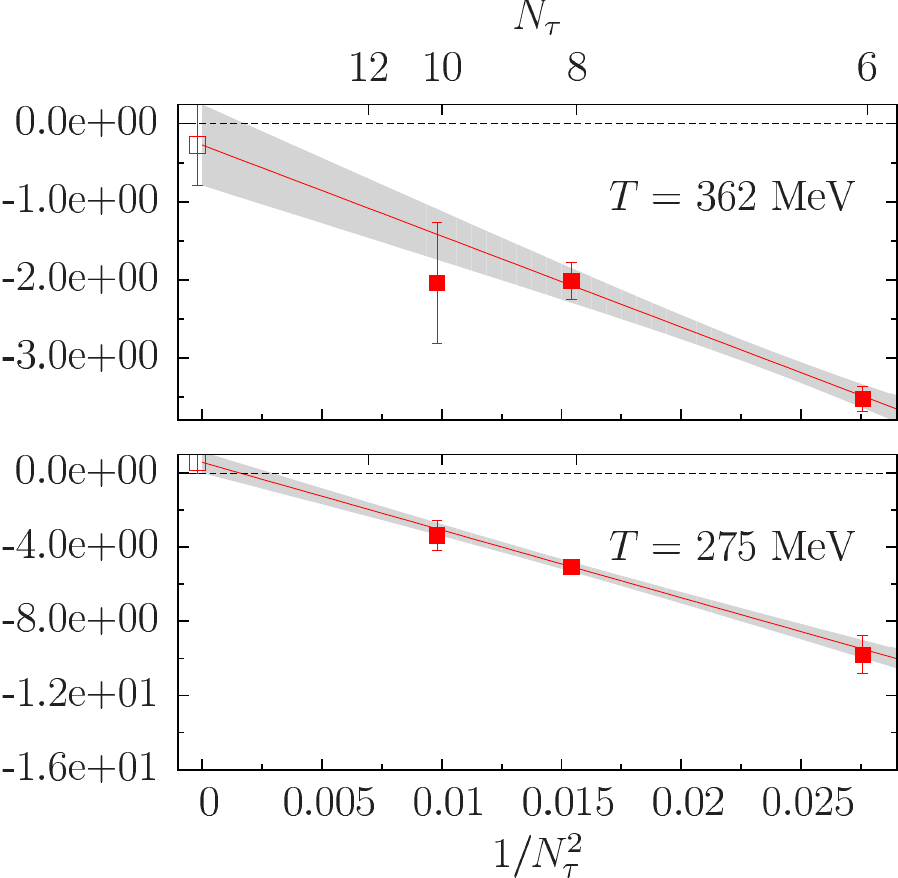}  
 \caption{\links Check of the continuum limit of the trace anomaly 
 contribution originating from the derivative \wrt the untwisted quark 
 mass $m$. We show data for the B mass ensemble at two values of the 
 temperature. 
 \mitte  The same for the C mass. 
 \rechts The same for the D mass.}
\label{fig_m0contlim}
\end{figure}

However, the $m$-derivative term in \Eq{eq:traceanomaly} 
containing $B_m$ does not contribute in the continuum limit. 
As will be shown in Appendix \ref{sec_m0symanzik} from a Symanzik expansion,
the subtracted vacuum expectation value of the operator arising from 
the $m$-derivative, $\vevsub{\sum_x \bar \chi_x \chi_x}$, is a pure lattice 
artefact at maximal twist and is vanishing in the continuum limit as 
$\mathcal{O}(a^2)$. We have checked this numerically by studying the
contribution to the trace anomaly from the term in question.  
\Fig{fig_m0contlim} shows the continuum limit of 
$B_m \vevsub{\sum_x \bar \chi_x \chi_x}$. As can be seen in the figure, 
the extrapolations to $1/N_\tau \to 0$ of this term are compatible with 
zero in all studied cases. Therefore, we have not included the contribution 
of this term upon evaluating \Eq{eq:traceanomaly} right from the beginning.

\begin{figure}[htb] 
\centering
  \includegraphics[width=0.5\textwidth]{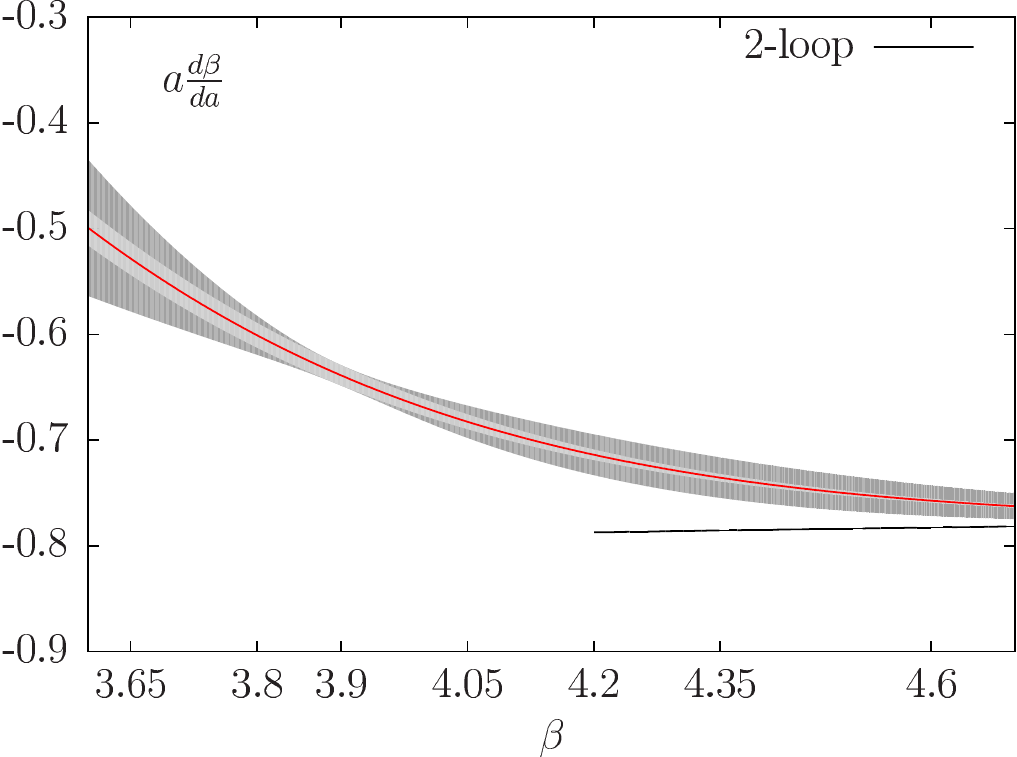}
 \caption{The $\beta$-function obtained according to \Eq{eq:betafunction} 
 from fitting expression (\ref{eq:betafunfit}) to the chirally extrapolated 
 data of the Sommer scale $\rchi$. We also show the perturbative 2-loop
 expectation at large couplings as obtained from \Eq{eq:twoloop}.}
\label{fig_betafun}
\end{figure}

The $B$-functions $B_\beta$ and $B_\mu$ in \Eq{eq:deri} are evaluated 
non-perturbatively from $T=0$ lattice data closely following 
\Cite{Cheng:2007jq}. In all cases we maintain the correct perturbative 
behavior of the $B$-functions and incorporate it explicitly into fit 
functions to $T=0$ data. The function $B_\beta$, directly related to the 
non-perturbative $\beta$-function and entering \Eq{eq:traceanomaly}
as a multiplicative factor, is evaluated by means of the following identity
in terms of the chirally extrapolated Sommer parameter $\rchi$:
\begin{equation}
    B_\beta = \left ( a \frac{d \beta}{d a} \right ) = 
             - \rchi \left ( \frac{d \rchi}{d \beta} \right )^{-1} \;.
    \label{eq:betafunction}
\end{equation} 
Using \Eq{eq:betafunction} and the two-loop expression \Eq{eq:twoloop}
we obtain the following asymptotic formula valid at large 
inverse squared
coupling $\beta$:
\begin{equation}
  B_\beta \left ( \beta \right ) 
   = - 12 \beta_0 - 72 \frac{\beta_1}{\beta} \;.
\label{eq:pert_Bfunction}
\end{equation}  
The interpolation of $B_\beta$ determined from the fit  
is shown in \Fig{fig_betafun} together with the 2-loop perturbative 
expectation according to \Eq{eq:pert_Bfunction}. The grey band in the graph
shows the error from the fit which is obtained by means of a bootstrap
analysis. The level of the error is of the order of 10 \% for low values of 
$\beta$ and goes down to the 3 \% level for higher values.

\begin{figure}[htb] 
\centering
\hfill
\includegraphics[width=0.45\textwidth]{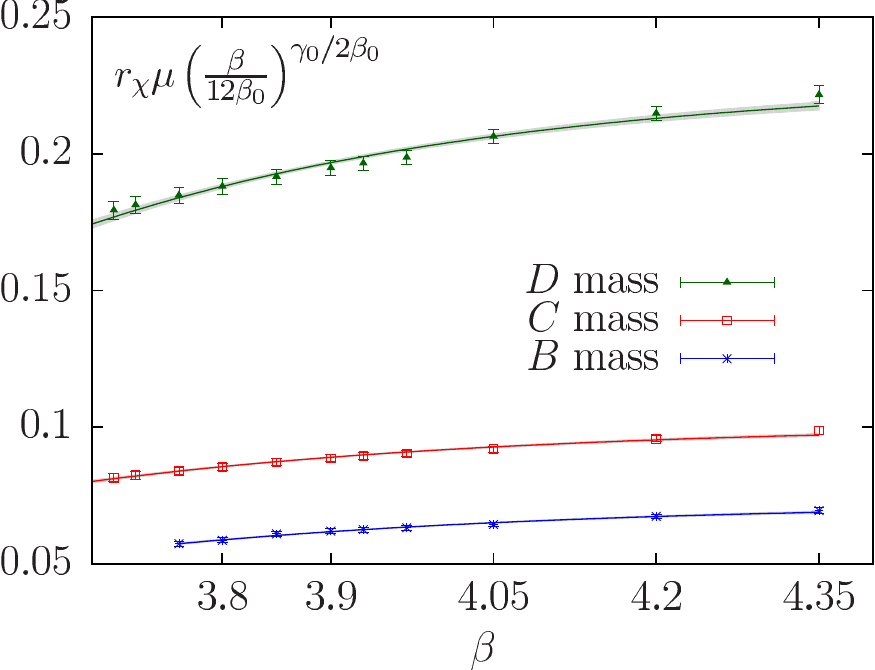}
\hfill
\includegraphics[width=0.45\textwidth]{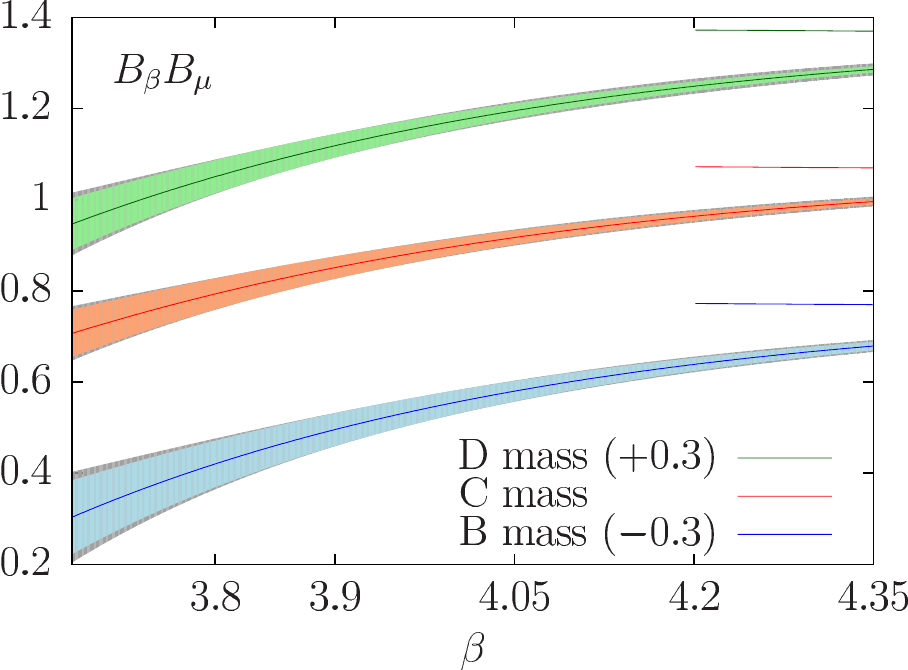}
\hfill
\caption{\links Fit of $r_\chi \mu$ with \Eq{eq:mufunctionfit} for 
all three masses. \rechts Combination of $B$-functions $B_\beta B\mu$ 
for all three masses. The perturbative asymptotic behavior is indicated 
by the lines at high values of the coupling. The curves for the B and D 
mass have been shifted for better visibility.}
\label{fig_mubeta}
\end{figure}

For the evaluation of the mass renormalization function $B_\mu$ we observe 
\begin{equation}
  \begin{split}
    B_\mu &= \frac{1}{(a \mu)} \frac{\partial (a \mu)}{\partial \beta} \\
          &= \frac{1}{(a \mu)} \left ( (a \mu) \frac{1}{a} 
       \frac{\partial a}{\partial \beta} 
     + \frac{1}{\rchi} \frac{\partial (r_\chi \mu)}{\partial \beta} \right )\\
          &= B^{-1}_\beta + \frac{1}{r_\chi \mu} 
       \frac{\partial (r_\chi \mu)}{\partial \beta} \,, \\
  \end{split}
\end{equation} 
where we have used the fact that $\frac{\partial r_\chi}{\partial \beta} = 0$, 
$r_\chi$ being the physical quantity that fixes the scale.
Accordingly we fit $(r_\chi \mu) (\beta)$ by the following expression
\begin{equation}
  r_\chi \mu = \left ( \frac{12 \beta_0}{\beta} \right )^{\gamma_0/2\beta_0} 
               P(\beta) \,.
\label{eq:mufunctionfit}
\end{equation}
The first factor gives the leading perturbative $\beta$-dependence of the 
mass (compare \eg \Cite{Chetyrkin:1999pq}) with $\gamma_0=1/(2 \pi^2)$. 
For the second factor we take a rational ansatz in terms of the ratio 
$R(\beta)$ as introduced in \Eq{eq:betafunfit},
\begin{equation}
  P(\beta) = a_\mu  \frac{ 1 + b_\mu R(\beta)^2}{1 + c_\mu R(\beta)^2}  \; .
\end{equation}
We employ our fit result of $\rchi(\beta)$ for building the product 
$(r_\chi \mu) (\beta)$. We have fixed $c_\mu \equiv 0$ for our main fits 
and take half the difference to fits with free $c_\mu$ but fixed 
$b_\mu \equiv 0$ into account as a systematic error. 
The fits for the three masses are shown in the left panel of 
\Fig{fig_mubeta}. We obtain reasonable fit results with 
$\chi^2/\rm{dof}= 0.26, 0.27, 0.59$ for the B, C and D mass, respectively. 
We show the result for the combination of $B$-functions $B_\beta B_\mu$ 
in the right panel of \Fig{fig_mubeta} and indicate the asymptotic behavior 
of this quantity ($B_\mu B_\beta = 1 + \frac{3}{\pi^3 \beta})$) at high 
values of the coupling. The colored shaded areas correspond to statistical 
errors and we have visualized the total errors including the systematic 
fit type related errors by grey bands. The error on $B_\beta$ has not been
included at this stage of the analysis. It is however accounted for
when computing the trace anomaly.

\section{Interpolation of $T=0$ Observables} \label{sec:t0int}

We have calculated the quantities needed in \Eq{eq:traceanomaly} 
for all pairs of values of $\beta$ and $(a \mu)$ that are available from 
ETMC \cite{Baron:2009wt}, see \Tab{tab_T0_ETM} in Appendix B.
Additionally we have substantially increased the $T=0$ data by additional
runs, see \Tab{tab_T0}. However, not every $T>0$ simulation point has 
been supplemented by an according $T=0$ simulation. Therefore, we have to
interpolate quantities entering \Eq{eq:traceanomaly} to the precise value 
of the twisted mass parameter $(a \mu(\beta))$ and the coupling $\beta$ 
that are used for the finite temperature runs. 

The mass dependence of the $T=0$ data points is fitted with cubic spline 
functions in the bare mass $a \mu$. We use the interpolated values from 
the fit. This interpolation is only necessary for some values of $\beta$ 
where the bare masses are not matched to the simulations at $T>0$ 
(in most cases at values of the couplings that have been studied by ETMC).

The interpolation of these (possibly ($a \mu$)-interpolated) values in 
the inverse coupling $\beta$ is performed using three types of fit functions 
in order to study the systematics corresponding to the choice of a specific 
fit function. Our first choice (further on called type A) is a simple 
polynomial function with varying degree $d_p$
\begin{equation}
  f^{d_p}_A(\beta) = \sum_{i=0}^{i = d_p} c_i \beta^{i} \,.
\end{equation}  
This ansatz may be extended by splitting the fit up into a low and a 
high $\beta$ part at a value of $\beta_{cut}$ leading to 
a fit type B
\begin{equation}  
  f_B(\beta) = \left\{
  \begin{array}{l l}
     & f_B^{low} = f^{d_p}_A(\beta) \quad \text{if $\beta < \beta_{cut}$}\\
     & f_B^{high} \quad \text{if $\beta > \beta_{cut}$}
  \end{array} \right.\,,
  \label{eq:ftb}
\end{equation} 
and we ensure smoothness of the function by an appropriately chosen 
$f_B^{high}$ 
\begin{equation}  
 f_B^{high}(\beta) = c_0 + c_1 (\beta-\beta_{cut}) 
   + \sum_{i=2}^{i = d_p} c_i \left ( \beta-\beta_{cut} \right )^{i} \,.
\end{equation}
As a third type of fit (type C) function we have considered again 
a cubic spline function.

\begin{table}
{
\small
\centering
     \begin{tabular}{c  |cc}
	$\chi^2/{\mathrm{dof}}$ & Type A  &  Type B\\
      \hline
         Ensemble B  & 1.7  &  1.5 \\
       \hline
         Ensemble C  & 2.2  &  1.6 \\ 
       \hline
         Ensemble D  & 3.4  &  1.5 \\          
    \end{tabular}
}
\caption{Fit quality results for $T=0$ interpolations of $\vev{S_g}$ 
providing the subtraction for the trace anomaly. 
For fit type A we show the best achieved $\chidof$ with $d_p = 5$ and 
for type B with $d_p = 4$ and varying $\beta_{cut}$, respectively.
}
\label{tab_tafit}
\end{table}

The central values for the $T=0$ subtraction of the gauge action contribution 
to the trace anomaly $\vev{S_g}$ are obtained from an average over three 
fits corresponding to the three fit types A, B and C discussed above.
For the former two the quality of the fit is indicated in \Tab{tab_tafit}. 
These had to be restricted to $\beta \ge 3.7$ for obtaining a good value 
for $\chidof$. We therefore restrict the analysis to $\beta \ge 3.7$ 
disregarding some simulated data at $\beta=3.65$ for the D and the C mass. 
For the B mass the fits have been conducted in the range 
$3.76 \le \beta \le 4.35$. In the case of fit type B we have included the 
best fit result (in terms of $\chidof$) obtained when varying $\beta_{cut}$ 
in \Eq{eq:ftb}. For type A we have restricted ourselves to $d_p = 5$  
since only for this choice the quality of the fit was reasonable.  
Our final value is obtained from an average over the three fits and 
taking half the maximal deviation of either of the three fits from the 
central value into account as a systematic error which is added to the 
statistical errors as obtained from fit type A in quadrature.  

Since the number of available $T=0$ points is very limited above 
$\beta=4.0$ we had to take special care in order to obtain a reliable
interpolation for the inverse coupling $\beta = 4.25$ for the B mass and 
$N_\tau=12$. In this case the precision at $T>0$ is good enough to see 
a $\sim 2 \sigma$ effect on the value of $I/T^4$ corresponding to 
$\beta=4.25$. To this end we have fitted the above fit functions of type A 
and B with lower number of parameters there and used these interpolations 
for the subtractions at the inverse coupling parameter values 
$\beta = 4.25$ and $\beta = 4.35$. For the C mass and $N_\tau=12$ with 
less statistics the effect is at the $\sim 1 \sigma$ level only and 
we stick to the analysis in terms of type A, B and C fits fitted globally 
to all values of the coupling.

\begin{figure}[htb] 
\centering
\hfill
\includegraphics[height=4.5cm]{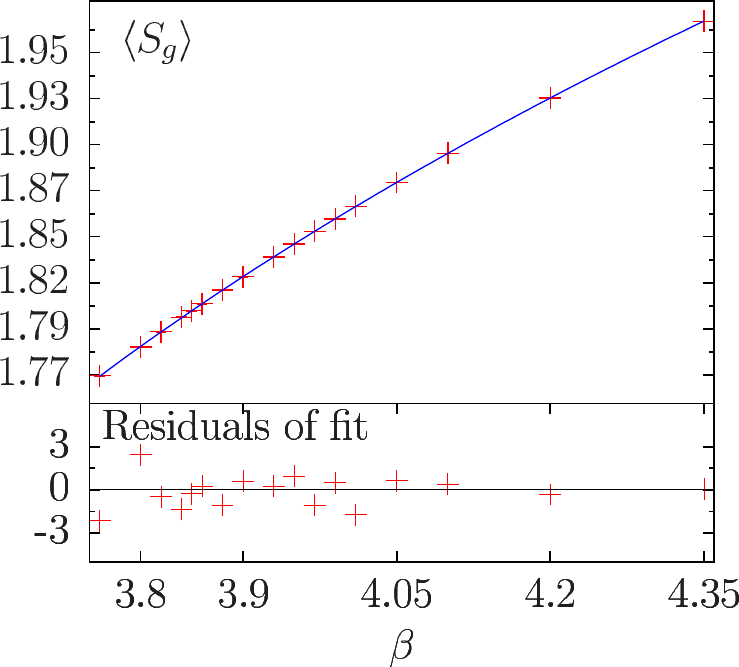} 
\hfill
\includegraphics[height=4.5cm]{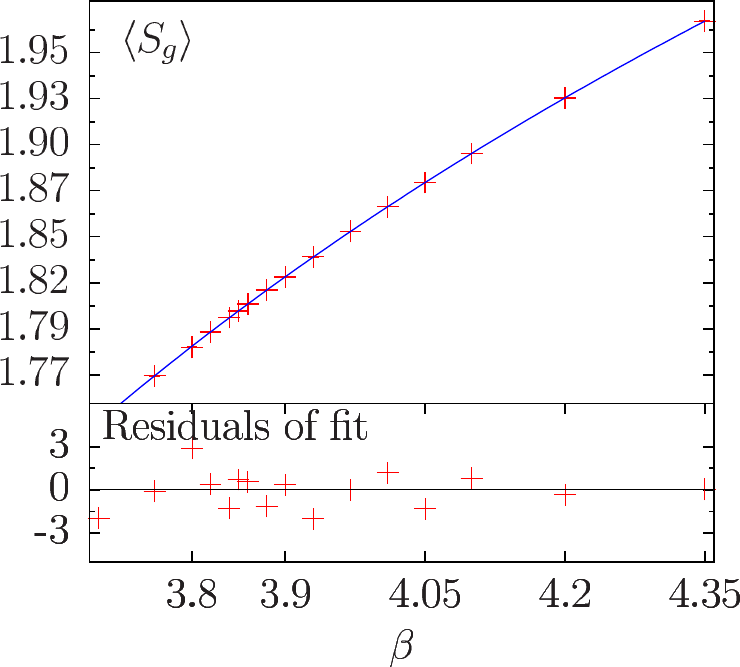}
\hfill
\includegraphics[height=4.5cm]{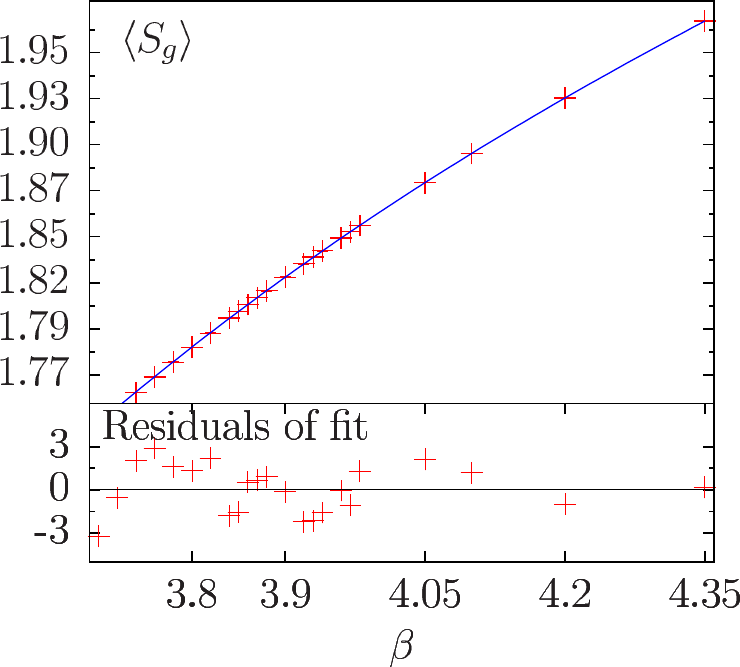}
 \caption{Interpolation of ($T=0$) values in $\beta$ 
for the gauge action contributions to the trace anomaly for 
(from left to right) the B, C and D mass. We show the outcome of a 
fit using fit type A with $d_p=5$. In each case we show the residuals 
of the fits in the lower panels in order to illustrate the quality of 
the interpolation. From left to right the resulting values for 
$\chi^2 / \mathrm{dof} $ have been $1.7$, $2.2$ and $3.4$, 
respectively.}
\label{fig_zeroTbeta}
\end{figure}

In \Fig{fig_zeroTbeta} we show results of the fits of type A for the 
three mass values. As from the figures themselves it is impossible to 
estimate the quality of the fit due to the small errors, we show the 
residuals of the fits (\ie the difference of the data and the fit 
normalised by the corresponding errors) in the lower panels of the 
figures. Having in mind the reasonable values of $\chi^2/{\mathrm{dof}}$
obtained we underline the fact that the different kinds of fits 
have provided curves which nicely fall on top of each other.

For the fermionic contribution $\vev{S_f}$ we used exclusively the
fit type C for subtracting the divergent contribution at $T=0$. 
The reason is that for the B ensemble the tuning of the mass $\mu$ has 
not been done on the same footing for all couplings. While in the near 
vicinity of the crossover the one-loop $\beta$-function has been used, 
we have opted for the two-loop $\beta$-function at larger as well as 
smaller couplings. For $N_\tau = 10$ the mass has been even tuned only
very approximately in the range $3.86 \le \beta \le 3.93$ and 
$a \mu = 0.006$ has been set. Since the divergence to be subtracted 
is of the form $\sim a\mu / a^3$, and thus sensitive to the mass, bad 
fit results can be expected when the precise value of $a \mu(\beta)$ 
is slightly changed when varying $\beta$. We note however that this 
slight variation in the way of tuning $a \mu$ does not affect the tuning 
of the physical pion mass, \ie the line of constant physics, as was 
shown in section \ref{sec:lcp}.

\section{Trace Anomaly Results} \label{sec:traceanomaly}
\begin{figure}[htb] 
\centering
\includegraphics[width=0.3\textwidth]{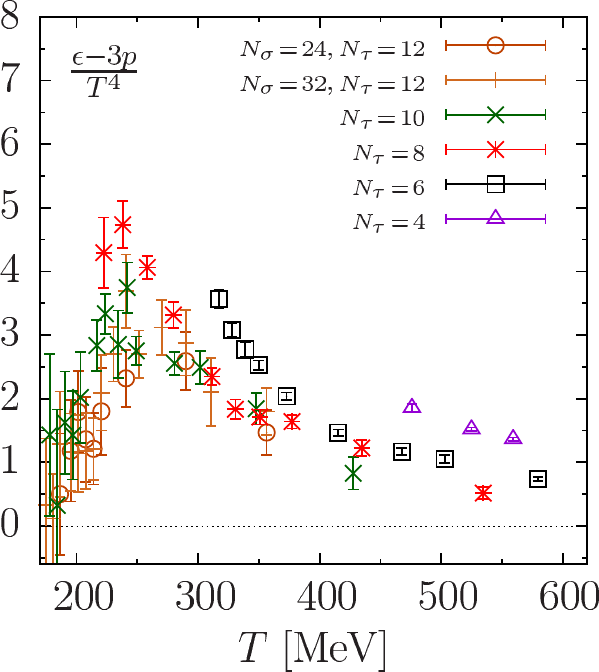}
\includegraphics[width=0.3\textwidth]{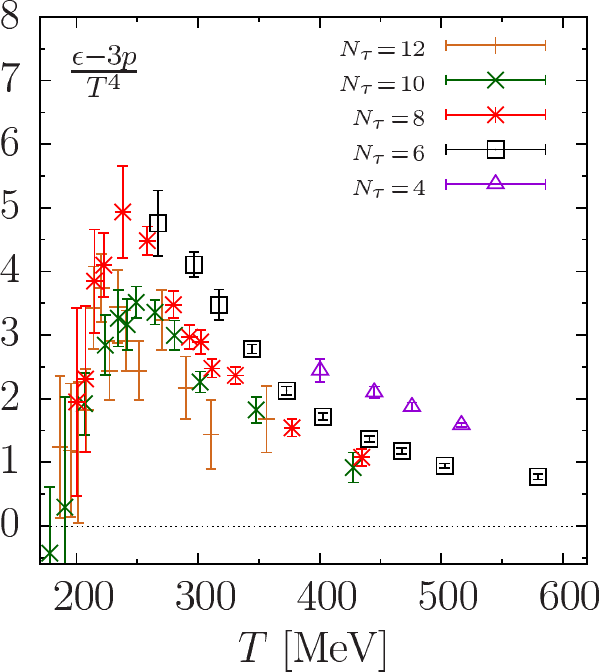}
\includegraphics[width=0.3\textwidth]{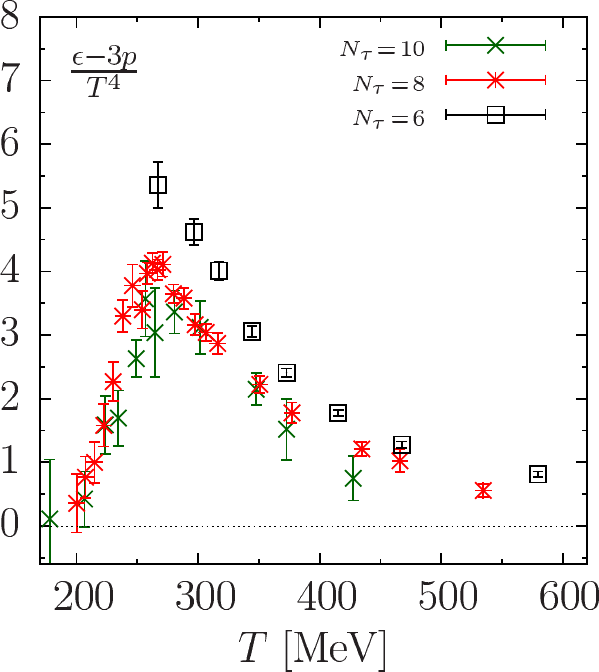}
\caption{\links The trace anomaly for the B mass obtained for different 
values of the temporal extent $N_\tau$. \mitte The same quantity for the 
C mass. \rechts The same quantity for the D mass. For the B mass the 
results obtained on the smaller spatial volume are superimposed slightly 
shifted for better visibility.}
\label{fig_B12em3p_nocorr}
\end{figure}

In this section we present our results for the trace anomaly and the therefrom
derived thermodynamic quantities for the B, C and D ensembles. The data is 
shown for varying $N_\tau$ in \Fig{fig_B12em3p_nocorr} where we observe 
severe lattice artefacts. In order to remedy this large effect we have studied 
what is known in literature as tree-level improvement \Cite{Borsanyi:2010cj}. 

The starting point for this method are the bosonic and fermionic pressures per
degree of freedom in the non-interacting limit which read 
$p_B = \frac{\pi^2}{90} T^4$ and $p_F = \frac{7}{8}\frac{\pi^2}{90} T^4$, 
respectively. Upon counting the number of bosonic and fermionic degrees of 
freedom for $N_f=2$ QCD we obtain the Stefan-Boltzmann limit of the pressure 
as
\begin{equation}
  \frac{p_{\mathrm{SB}}}{T^4} = 
  \left(16 + \frac{7}{8} \times 24 \right) \frac{\pi^2}{90} \approx 4.0575\,.
 \label{eq:pSB}
\end{equation}

On the lattice the free limit pressure $p_{\mathrm{SB}}^L$ receives 
$N_\tau$-de\-pen\-dent corrections that vanish in the continuum limit.
It has been calculated for the twisted mass action in \Cite{Philipsen:2008gq}.
Through the mass dependence of the fermion propagator, $p_{\mathrm{SB}}^L$ 
as well as $p_{\mathrm{SB}}$ depend in general on the ratio $\frac{m_R}{T}$
of renormalised quark mass and temperature.
However, this dependence is weak, the change being of the order of below 
1 \% when varying $m_R$ in the ranges of the twisted masses we have simulated. 
We have used the ratio $p_{\mathrm{SB}}^L/p_{\mathrm{SB}}$ to correct the 
trace anomaly data for its leading cutoff effects. 

Together with the corresponding ratio for the tree-level Symanzik improved 
gauge action that can be found  in \Cite{Karsch:2000ps}%
\footnote{Since the ratio rapidly approaches unity for increasing $N_\tau$ 
we adopt a value of $1$ for $N_\tau = 12$ which induces a negligible error.} 
we obtain the following correction factors that are used throughout 
this work:
\begin{center}
     \begin{tabular}{c | ccccc}
        $N_\tau$ 		   & 4 & 6 & 8 & 10 & 12 \\
        \hline \hline
        $p_{\mathrm{SB}}^L/p_{\mathrm{SB}}$  & 2.576 & 1.631 & 1.263 & 1.134 & 1.082
     \end{tabular} \; .
\end{center}
The tree-level correction of the trace anomaly then amounts to making the
 following replacement in the whole temperature interval:
\begin{equation}
 \left ( \frac{I}{T^4} \right ) \Rightarrow \left ( \frac{I}{T^4} \right ) 
    \Big{/} \left ( \frac{p_{\mathrm{SB}}^L}{p_{\mathrm{SB}}} \right )\,.
\label{eq:pressurecorrection}
\end{equation}
\begin{figure}[htb] 
\centering
\includegraphics[scale=0.55]{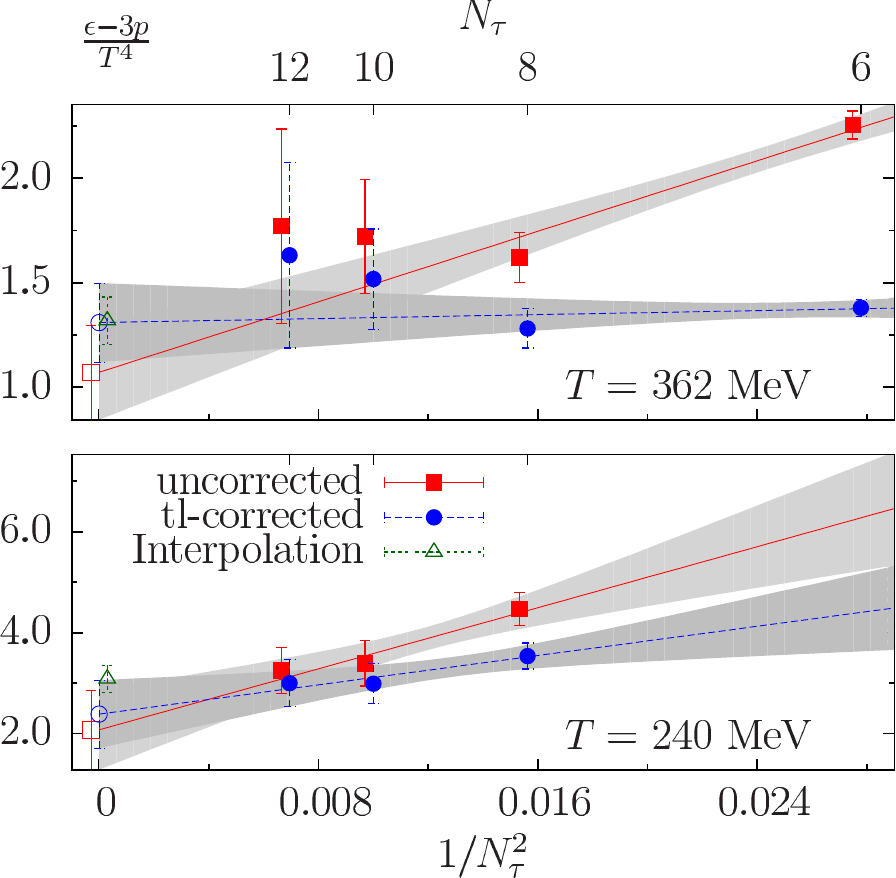}
\includegraphics[scale=0.55]{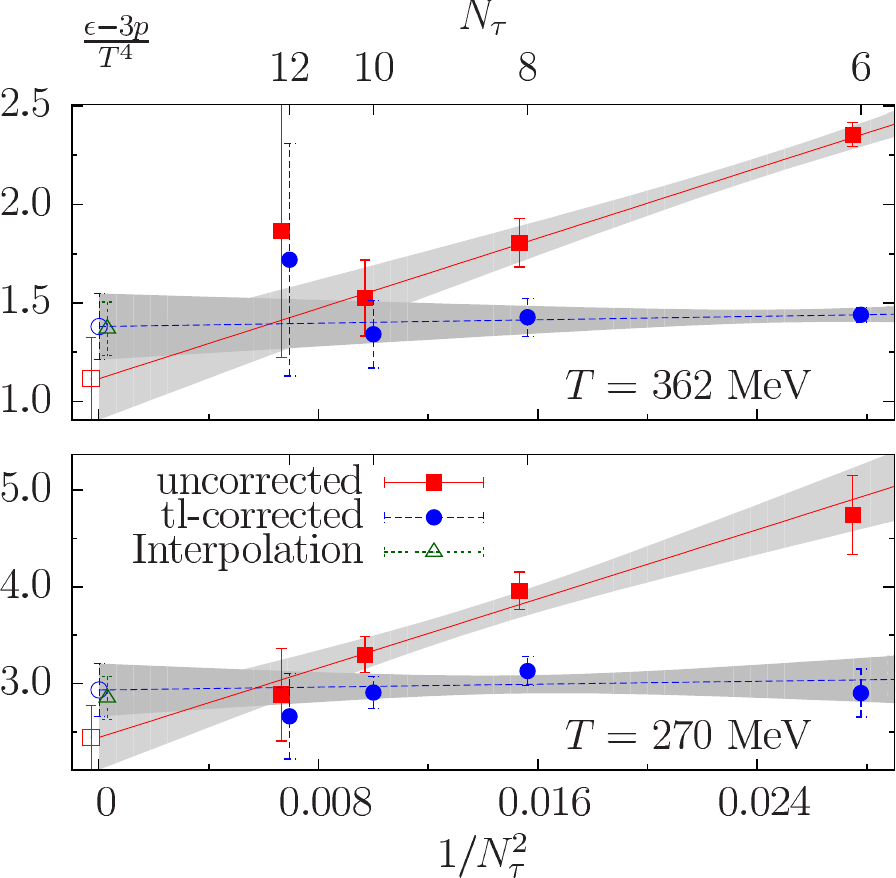}
\includegraphics[scale=0.55]{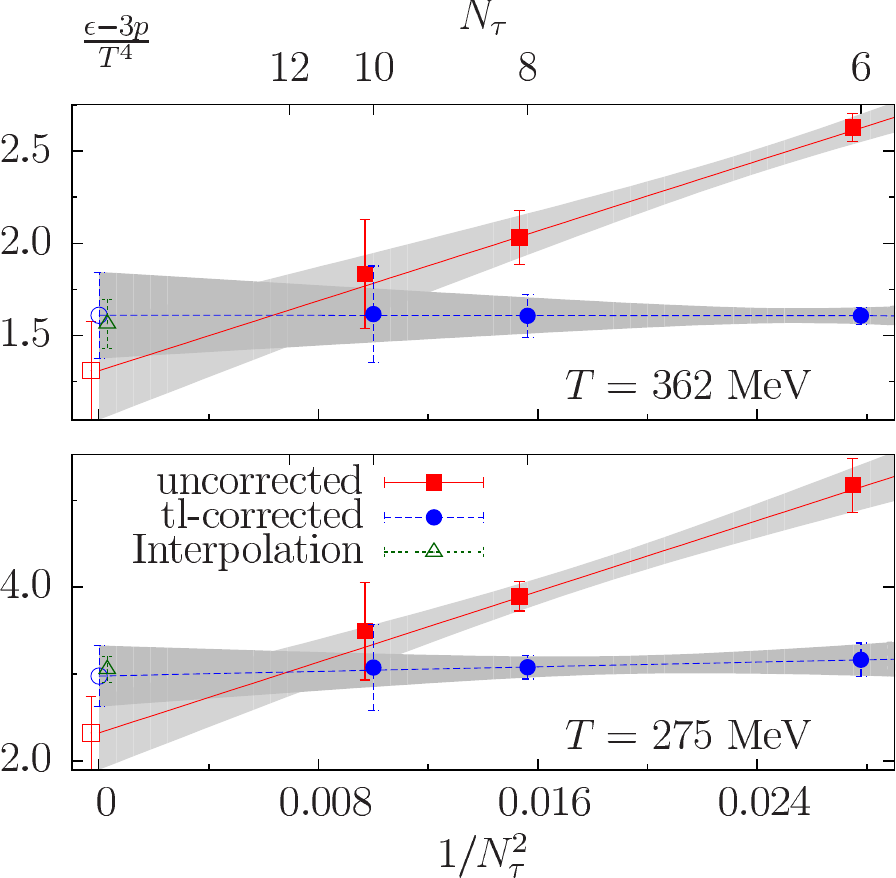}
\caption{Continuum limit of the trace anomaly for three masses and 
for in each case two values of the temperature once with tree-level 
correction (blue circles and lines) and once without (red squares and 
lines). We compare the continuum limit results for the two extrapolations 
with the continuum estimate provided by the global fit \Eq{eq:em3p_interpol} 
at the same temperature (green triangles).}
\label{fig_em3p_contlim}
\end{figure}

The tree-level correction of the trace anomaly may be checked by studying 
the continuum limit of $I/T^4$ with and without the correction in place.
To this purpose we show in \Fig{fig_em3p_contlim} a comparison of 
two different ways to take the continuum limit of the trace anomaly for 
the three ensembles. In each case we consider two values of 
the temperature $T = (240$ and $362) \mathrm{~MeV}$, 
$T = (270$ and $362) \mathrm{~MeV}$ and $T = (275$ and $362) \mathrm{~MeV}$ 
for the B, C and D ensemble, respectively.
The smaller temperature was chosen in the range of the maximum of 
the interaction measure, while the higher temperature is situated in the 
falling (right) flank. Data for different $N_\tau$ was interpolated  
using a second order polynomial fitted to the four data points closest 
to the given temperature under investigation. We perform continuum 
extrapolations linear in $1/N_\tau^2$ including $N_\tau = 12, 10, 8$ 
(where possible also $N_\tau=6$) once with the multiplicative correction 
(\Eq{eq:pressurecorrection}) in place and once without it. We observe that 
both procedures lead to compatible continuum limit values matching each 
other within two standard deviations for the trace anomaly. The correction 
leads in general to a flatter continuum limit than we observe for the 
uncorrected data. Apart from $T=240$~MeV for the B mass (where no 
$N_\tau=6$ data point is available) the corrected results are even compatible 
with a flat continuum limit. Moreover, the corrected trace anomaly at the two 
largest temporal extents ($N_\tau = 8,10$ for the D mass and 
$N_\tau = 10, 12$ for the B mass) are in all cases compatible 
with each other within errors. 

\begin{figure}[htb] 
\centering
\includegraphics[width=0.3\textwidth]{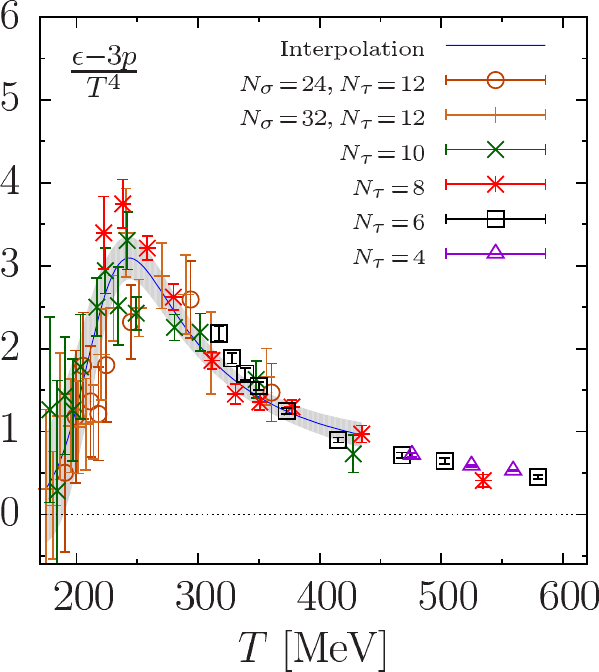}
\includegraphics[width=0.3\textwidth]{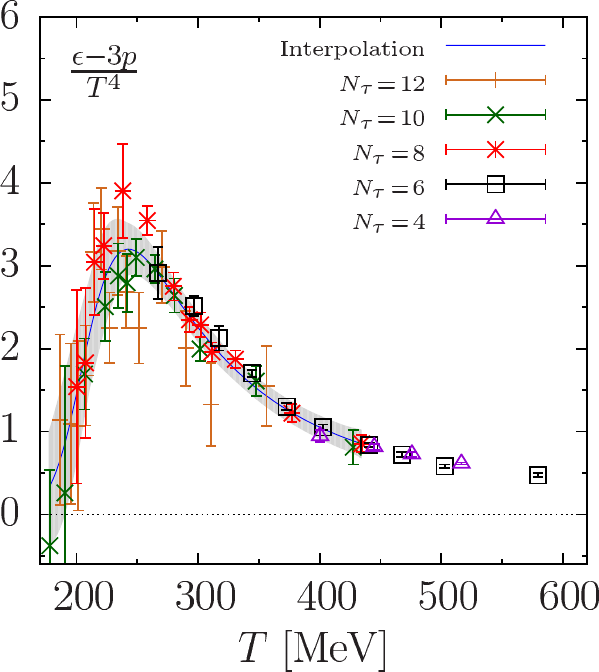}
\includegraphics[width=0.3\textwidth]{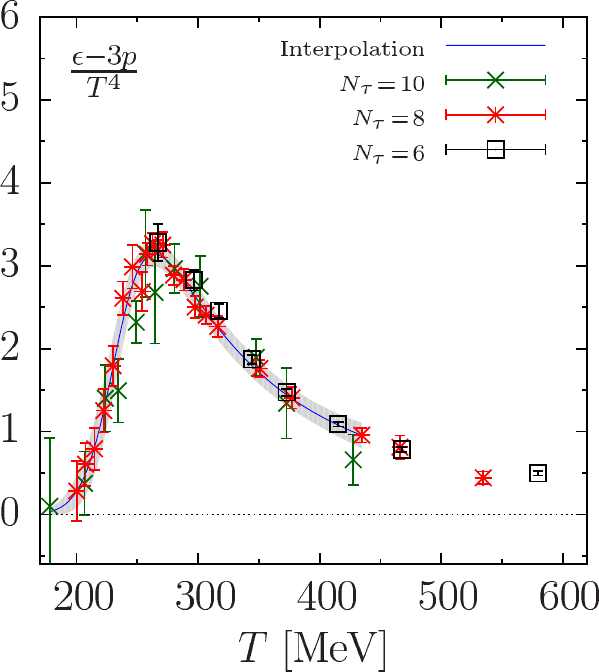}
\caption{\links The tree-level corrected trace anomaly for the B mass 
obtained for different values of the temporal extent $N_\tau$. 
$T_\chi$ and $T_\mathrm{deconf}$ are located at $193$ and $219$ MeV, 
respectively. \mitte The same quantity for the C mass with $T_\chi$ and 
$T_\mathrm{deconf}$ located at $208$ and $225$ MeV, respectively.
\rechts The same quantity for the D mass with $T_\chi$ and 
$T_\mathrm{deconf}$ located at $229$ and $244$ MeV, respectively.
Also shown is the result of a combined fit of the interpolation formula 
\Eq{eq:em3p_interpol} to the $N_\tau = 8,10$ and $12$ data 
($N_\tau = 6,8$ and $10$ in case of the D mass). For the B mass the 
results obtained on the smaller spatial volume are superimposed slightly 
shifted for better visibility. This data however has not been included 
in the fit.}
\label{fig_B12em3p}
\end{figure}

The integration of \Eq{eq:pressureintegral} is performed by fitting a 
modified version of an ansatz used in \Cite{Borsanyi:2010cj} to the 
available lattice data for $\frac{I}{T^4}$ (discarding those for 
$N_\tau = 4$).
\begin{equation}
  \frac{I}{T^4} = \left ( 1 + \frac{a_2}{N_\tau^2} \right ) \times 
  \exp{\left ( - h_1 \bar t - h_2 {\bar t}^2 \right )} \cdot
  \left ( h_0 + \frac{f_0 \left \{ \tanh{\left ( f_1 \bar t + 
  f_2 \right )} \right \} }{1 + g_1 \bar t + g_2 {\bar t}^2 } \right) \,.
  \label{eq:em3p_interpol}
\end{equation}
While in \Cite{Borsanyi:2010cj} the normalisation temperature $T_0$ 
in the dimensionless ratio $\bar t \equiv T/T_0$ has been fixed we 
let $T_0$ vary in the fit. Since we observe large cutoff effects in 
our trace anomaly results we furthermore include a multiplicative correction
term into our fit function incorporating the leading and sub-leading 
$N_\tau$ dependence.  

Table \ref{tab_em3pfit} lists the best fit parameters together with 
errors of fits of the interpolation formula \Eq{eq:em3p_interpol} to
the trace anomaly data for the B, C, and D ensembles, respectively,
after the trace anomaly has been corrected using the tree-level correction 
\Eq{eq:pressurecorrection}.
The table also provides the $\chi^2/\mathrm{dof}$ values of these fits.
The interpolation curves are illustrated together with the data corresponding 
to the three pion mass values in \Fig{fig_B12em3p}.

The error of the interpolation indicated by a grey band in these figures 
is evaluated as follows. From fits to bootstrap samples of our data we 
estimate a first error of our interpolation, giving rise to the errors on 
the fit parameters presented in \Tab{tab_em3pfit}. A second error is obtained 
by fitting the interpolation function to the data shifted by one standard
deviation in the upper and lower directions and measuring the deviation to 
the fit of the original data. Both errors are then added in quadrature. 
We have adopted this rather non-standard method because we have observed 
that the first of these errors (originating from the bootstrap analysis) 
is very small as compared to the uncertainties of the data themselves. 
This is especially true for the low temperature region. Thus by considering 
the pure fit error we would certainly have underestimated 
the error of the trace anomaly interpolation there.

For the D ensemble we have included $N_\tau = 10$ and $N_\tau = 8$
into the fit, while for the C and B ensemble we fit $N_\tau = 12,10$ and $8$.
This approach, which assumes a behavior constant in $N_\tau$ towards 
the continuum limit, is justified by the effectiveness and reliablility of 
the tree-level correction. The latter effectively superimposes data from 
different $N_\tau$ as can be seen from \Fig{fig_B12em3p}. We have 
explicitly checked the outcome of our global fits for two values of the 
temperature in \Fig{fig_em3p_contlim}, where we compare standard continuum
extrapolations in $1/N_\tau^2$ for data with and without tree-level correction 
to the continuum estimate provided by this fit. In all cases we find 
compatible continuum results. 

\begin{table}
\centering
     \begin{tabular}{c |ccccc}
        \textsc{Ensemble} &  \multicolumn{5}{c}{\textsc{Parameters}} \\ 
        \hline
        \multirow{4}{*}{B} & $h_0$ & $h_1$ & $h_2$ & $f_0$ & $f_1$ \\ 
        \cline{2-6}
          & 0.20(13) &  -4.4(1.4) & 4.9(1.8) &  0.074(21) & 0.9090(3)\\
        \cline{2-6}
           & $f_2$ & $g_1$ & $g_2$ & $T_0$ &$\chi^2/\mathrm{dof}$\\
        \cline{2-6}
           &  5.5112(3) & -1.83(8) & 0.88(7) & 211(4) & 1.7 \\
        \hline 
        \hline          
        \textsc{Ensemble}  & \multicolumn{5}{c}{\textsc{Parameters}} \\ 
        \hline        
        \multirow{4}{*}{C} & $h_0$ & $h_1$ & $h_2$ & $f_0$ & $f_1$ \\ 
        \cline{2-6}
          & 0.03(3) & -8.7(2.5) & 6.8(2.8) & 0.021(9) &  0(3) \\
        \cline{2-6}
           & $f_2$ & $g_1$ & $g_2$ & $T_0$ &$\chi^2/\mathrm{dof}$\\
        \cline{2-6}
           & 1(4)  & -2.2(2) & 1.29(18) & 238(2) & 1.2\\
        \hline 
        \hline
        \textsc{Ensemble}  & \multicolumn{5}{c}{\textsc{Parameters}} \\ 
        \hline        
        \multirow{4}{*}{D} & $h_0$ & $h_1$ & $h_2$ & $f_0$ & $f_1$ \\ 
        \cline{2-6}
          & 0.05(7) & -5.3(7) & 5.3(8) & 0.09(2) & 1.15421(9)  \\
        \cline{2-6}
           & $f_2$ & $g_1$ & $g_2$ & $T_0$ &$\chi^2/\mathrm{dof}$\\
        \cline{2-6}
           & 5.73995(7) & -2.2(6) & 1.27(6) & 268(2) & 0.90\\
   \end{tabular}
\caption{Fit parameters obtained from fits of \Eq{eq:em3p_interpol} 
to the tree-level corrected trace anomaly data of the 
B, C, and D mass ensembles, respectively.}
\label{tab_em3pfit}
\end{table}

We conclude this paragraph with a discussion of finite size effects.
At $T>0$ the thermodynamic limit has been studied for the smallest 
pion mass. We have evaluated the trace anomaly for $N_\tau=12$ reducing the
spatial extent from $N_\sigma = 32$ to $N_\sigma = 24$. As can be seen 
from \Fig{fig_B12em3p} the results on the smaller volume are compatible 
(within the large errors) with the result obtained in the larger volume.

\section{Pressure and Energy Density} \label{sec:thermo}

From the fitted interpolation of the interaction measure \Eq{eq:em3p_interpol} 
it is straightforward to calculate the pressure by performing a numerical 
integration starting in all cases from the lowest available data point of 
$I/T^4$ where we set the pressure equal to zero. In other words we set 
$p_0 = 0$ in \Eq{eq:pressureintegral} with $T_0$ being our smallest 
temperature $T=174~\mathrm{MeV}$ ($T=177~\mathrm{MeV}$) for the B (C and D) 
mass. In section \ref{sec:hrg} we try to estimate $p_0$ from a comparison 
of our $N_f=2$ lattice data at unphysically high masses to adapted HRG models.

\begin{figure}[htb] 
\centering
\includegraphics[width=0.3\textwidth]{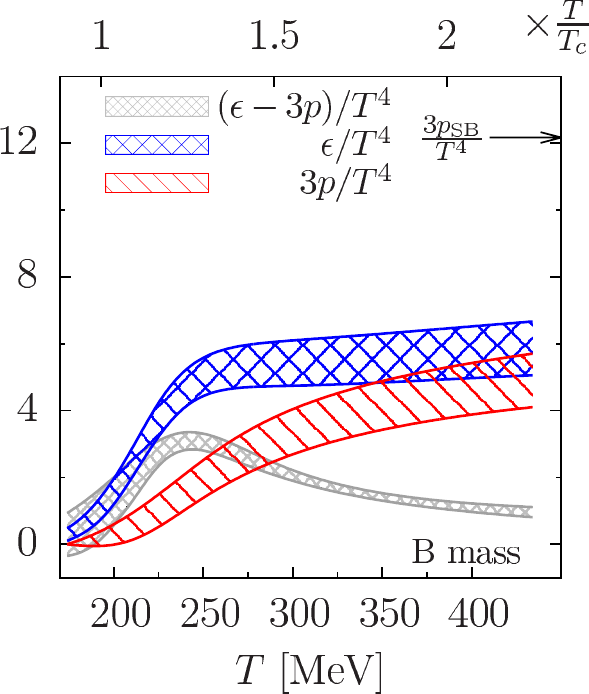}
\includegraphics[width=0.3\textwidth]{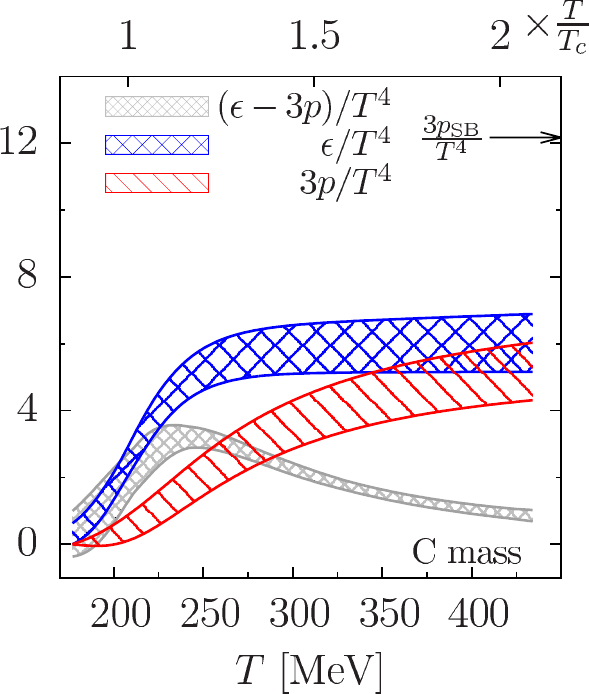}
\includegraphics[width=0.3\textwidth]{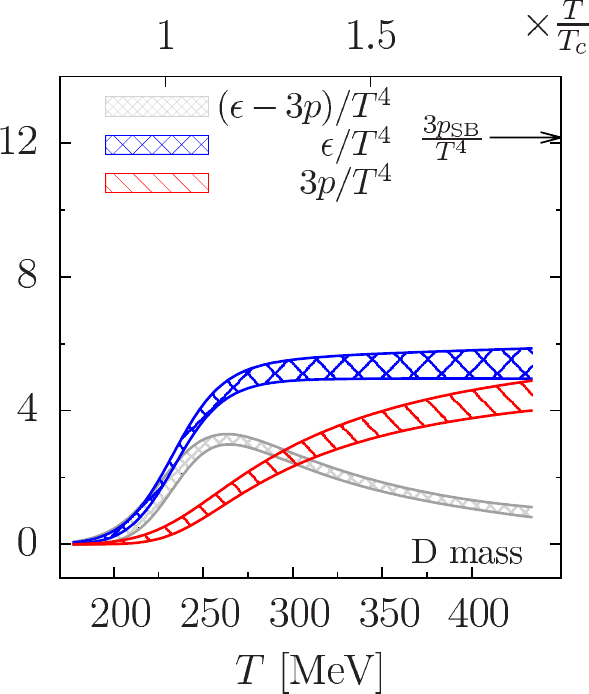}
\caption{\links Final result for the pressure $p$ and the energy density 
$\epsilon$ in units of $T^4$ for the B mass ensemble. We also show once 
more the interpolation of the trace anomaly used for integrating the pressure. 
The arrow in the upper right corner indicates the expected Stefan-Boltzmann 
limit for the pressure. On top of the panels we provide the temperature in 
units of $T_c \equiv T_\chi$. \mitte The same for the C mass. 
\rechts The same for the D mass.}
\label{fig_3p_epsilon}
\end{figure}

In this way we obtain the pressure (and the energy density from adding 
three times the pressure to $I$) for all temperatures in the temperature 
interval covered by our simulations. We do not restrict ourselves to the 
points in $T$ where we actually have lattice data, but rather give the
corresponding error channels for all upper integration bounds. This seems to 
us the most natural choice as we have included into the interpolation fits 
to $I/T^4$ data from several values of $N_\tau$.
In \Fig{fig_3p_epsilon} we show our results for the pressure ($3 p$) as well 
as the energy density ($\epsilon$) as a function of the temperature. 
At the top of the three panels in this figure we also mark the temperature
in units of the pseudo-critical temperature $T_c \equiv T_\chi$ the latter 
determined from the maximum of the chiral susceptibility in each mass case.
We used the estimate originating from the largest $N_\tau$ in all cases.

The energy density features a sharp rise around $T_c$ signalling the 
transition into the quark-gluon plasma regime. At temperatures of about
$\sim 1.3 ~T_c$ however, the increase has stopped and we observe an 
almost constant behavior up to the largest temperatures considered.
This feature is also observed by other groups, \cf 
\Cite{Cheng:2009zi, Borsanyi:2010cj,AliKhan:2001ek}.
At large temperatures we can confront pressure and energy density 
to the ideal gas Stefan-Boltzmann pressure (\Eq{eq:pSB}),
which is indicated by the black arrow to the right of the figures. 
At our largest accessible temperature corresponding to $T/T_c \sim 2$ our 
computed energy density attains roughly half of its expected asymptotic 
Stefan-Boltzmann limit value.

Another observable of phenomenological interest derivable from the basic 
bulk thermodynamic quantities $p$ and $\epsilon$ is the velocity of
sound of the hot medium which is defined as the derivative 
of the pressure with respect to the energy density 
\begin{equation}
  c_s^2 = \frac{d p}{d \epsilon} 
\end{equation}
and may be calculated from the ratio $p/\epsilon$ by means of
the following identity \cite{Cheng:2009zi}:
\begin{equation}
  \frac{d p}{d \epsilon} = \epsilon \frac{d (p / \epsilon)}{d \epsilon} 
                           + \frac{p}{\epsilon} \; .
\label{eq:sos_calc}
\end{equation}

\begin{figure}[htb] 
\centering
\includegraphics[width=0.3\textwidth]{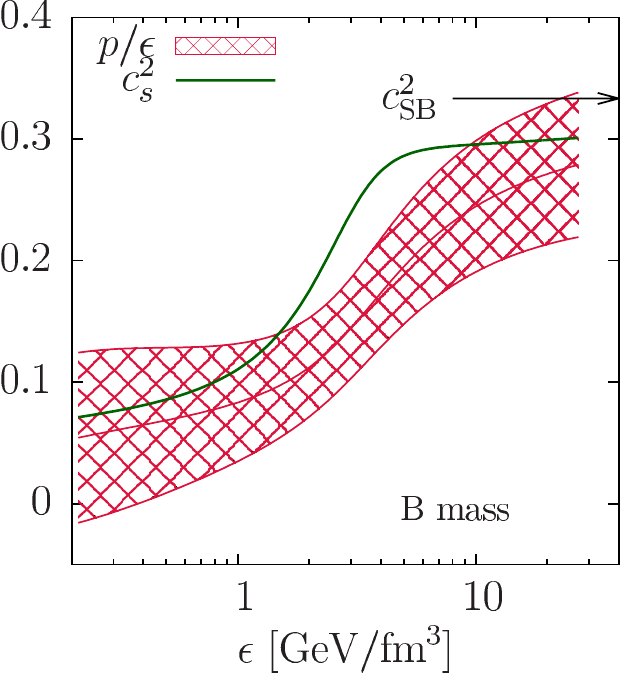}
\includegraphics[width=0.3\textwidth]{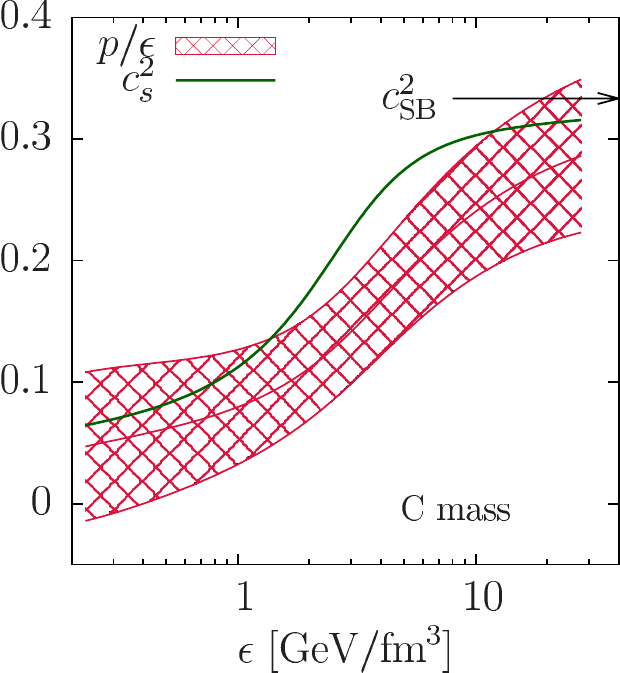}
\includegraphics[width=0.3\textwidth]{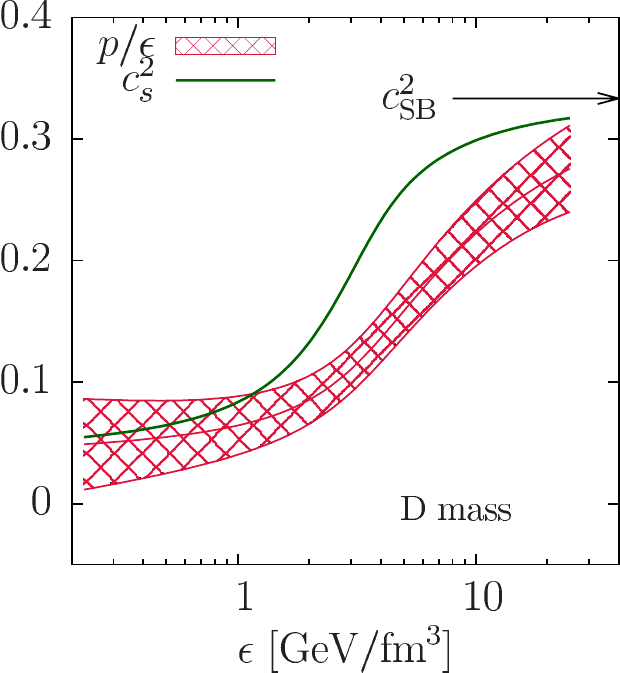}
\caption{\links The ratio $p/\epsilon$ for the B ensemble as a 
function of the energy density in units of $\mathrm{GeV} / \mathrm{fm}^3$. 
We also show the speed of sound squared $c_s^2$ obtained from 
$p/\epsilon$. Arrows indicate the expected large $T$ Stefan-Boltzmann
 limit given by $1/3$. \mitte The same for the C mass. 
\rechts The same for the D mass.}
\label{fig_sos_epsilon}
\end{figure}

In \Fig{fig_sos_epsilon} we show our result for the ratio of pressure and 
energy density as well as the speed of sound $c_s^2$ as a function of the 
energy density in units of $\mathrm{GeV} / \mathrm{fm}^3$.
The ratio $p/\epsilon$ is evaluated most directly from $p$ and $\epsilon$,
whereas the speed of sound $c_s^2$ is evaluated according to \Eq{eq:sos_calc}
from its derivative. We do not calculate any error so far for the velocity 
of sound, as the error on the basic quantity $p/\epsilon$ itself is already 
very large. At large temperatures we observe that the limiting 
Stefan-Boltzmann value of $\left ( p/\epsilon \right )_\mathrm{SB} = 1/3$ 
is nicely approached. However, we are not able at the current precision to 
resolve the dip at small temperatures that is observed in $p/\epsilon$  
results from staggered simulations \cite{Borsanyi:2010cj, Bazavov:2011nk}.       

\section{Hadron Resonance Gas Model: fixing the integration constant $p_0$} 
\label{sec:hrg}

In the hadronic phase at temperatures below the crossover transition hadrons 
and resonances form the relevant degrees of freedom that may be thermally n
excited. It has been argued, that in this region of temperature a gas of free, 
non-interacting hadrons and resonances could provide a good approximation 
to the interacting thermal medium. A comparison of the hadron resonance 
gas (HRG) model with lattice data has been conducted for instance in 
\Cite{Borsanyi:2013bia, Bazavov:2014pvz}. Good agreement with results of 
non-perturbative lattice evaluation is found for various quantities even up 
to the crossover temperature. 

\begin{figure}[htb] 
\centering
\includegraphics[width=0.45\textwidth]{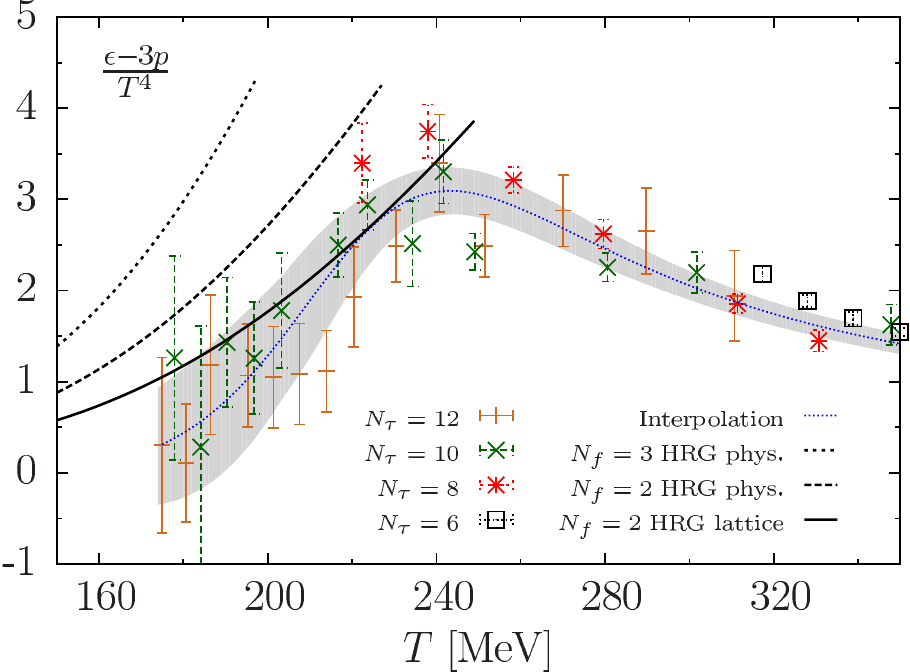}
\includegraphics[width=0.45\textwidth]{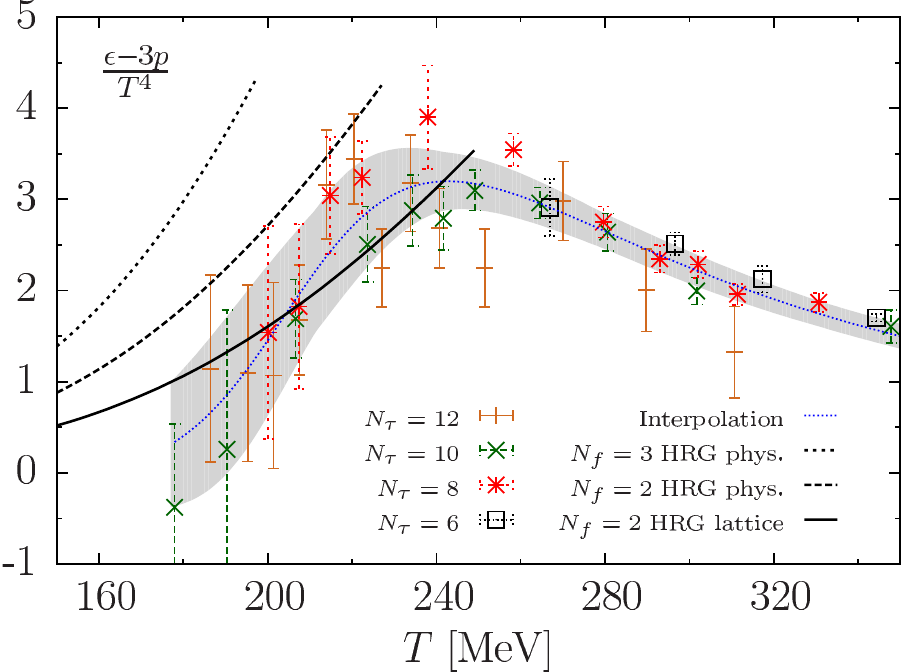}
\caption{Comparison of the interaction measure in the low temperature region 
to the predictions of several HRG model adaptations. \links The data and 
analysis corresponding to the B mass is shown. \rechts The same for the 
C mass. See text for details.}
\label{fig_em3p_hrg}
\end{figure}

At vanishing chemical potential the free pressures of mesons ($M$) and 
baryons ($B$) can be written as 
\begin{equation}
 \frac{p^{\mathrm{HRG}}}{T^4} =  
 \frac{1}{V T^3} \sum_{i \in \mathrm{Mesons}} \ln{Z^M_{m_i}}(T,V) + n
 \frac{1}{V T^3} \sum_{i \in \mathrm{Baryons}} \ln{Z^B_{m_i}}(T,V)\,,
\label{eq:hrg_pressure}
\end{equation}
where according to Bose- and Fermi- statistics for mesons and baryons,
respectively, and in terms of the energies 
$E_i(m_i, k) = \sqrt{(m_i^2+k^2)}$ and state degeneracies $d_i$ 
the contributions $\ln{Z^{M/B}_{m_i}}$ are given as
\begin{equation}
 \ln{Z^{M/B}_{m_i}} = d_i \frac{V}{2 \pi^2} \int_0^\infty k^2 
 \ln{ \left (1 \mp e^{-E_i(m_i, k)/T} \right )} dk \,.
\end{equation}
We have conducted a three-fold analysis confronting our lattice data 
of the interaction measure for the B and the C mass with the interaction 
measure provided by above formula upon taking the derivative \wrt 
temperature as prescribed in \Eq{eq:pressurederiv}. 
There are three options. Firstly we may include all known physical 
states as referenced by the PDG \cite{Agashe:2014kda} up to a certain 
cutoff mass, which we have set to $m_{cut} = 1.9 ~ \mathrm{ GeV}$ 
throughout. The heaviest meson we include is the $\pi_2(1880)$ and the 
heaviest baryon is the $\Delta(1905)$. As can be seen from 
\Fig{fig_em3p_hrg} the interaction measure evaluated with all physical 
states (corresponding to the $N_f=3$ curve in the pictures) overshoots 
the data at small temperatures significantly.

As another choice we may restrict the set of states entering 
\Eq{eq:hrg_pressure} to the ones with $S=0$, \ie to states without 
strangeness. Doing so already closes more than half of the gap between 
our lattice results and the HRG estimate from $N_f=2$ which is labelled 
``$N_f=2$ HRG phys.'' in the pictures. Since the value of pion masses 
we have considered in this work is yet somewhat above the physical 
value the remaining difference to a $N_f=2$ HRG model at physical 
masses is not unexpected. 
Along the ideas of \Cite{Huovinen:2009yb} we have therefore conducted 
a third approach and have used where possible the measured lattice 
mass spectrum data corresponding to unphysical pion masses obtained 
within the ETMC. We summarize in the list given below 
the mass information $M_B$ and $M_C$ for the states we have included 
in this analysis for the cases of the B and C ensembles and give the 
reference, where it has been published:\\
\begin{center}
\begin{tabular}{c|c|c|c}
 State & Reference & $M_B$ [GeV] &  $M_C$ [GeV] \\
 \hline
 $\rho$ 	&\cite{Jansen:2009hr}		& 0.943	& 0.858	\\
 $a_0$		&''				& 1.116	& 1.252	\\
 $b_1$		&''				& 1.603	& 1.529	\\
 $\eta_2$	&\cite{Jansen:2008wv}		& 1.008	& 1.066	\\
 $N$		&\cite{Alexandrou:2009qu}	& 1.209	& 1.282	\\
 $\Delta$	&''	 			& 1.517	& 1.589	\\ 
\end{tabular} 
\end{center}

In all cases at least two lattice spacings (corresponding to $\beta=3.9$ 
and $\beta=4.05$) as well as several values of the bare quark mass 
$a \mu$ have been studied. 
A detailed continuum extrapolation including three or more values of 
the cutoff would go beyond our possibilities in most of the cases.
For converting into physical units we have used the lattice 
spacings obtained in \Cite{Alexandrou:2010hf}, $a = 0.089$ and 
$a=0.070$ for $\beta=3.9$ and $\beta=4.05$, respectively.
Since in all cases cutoff effects are small, we have used in our analysis 
the value obtained from a spline interpolation of the hadron mass in
lattice units as a function of the bare quark mass at the finer lattice 
spacing and converted to physical units. The values are shown in the 
table above. 

As the mass splitting of different isospin states induced by the twisted 
mass term is mostly small (with the exception of the neutral pion !) 
and since no cutoff effects are visible in our trace anomaly data at low
temperature, we have neglected this splitting.  

Furthermore, in addition to the table above we could use the $\omega-\rho$ 
mass-splitting of $27~\mathrm{MeV}$ calculated in \Cite{McNeile:2009mx} 
which then fixed the $\omega$-mass in our analysis. 
The excited states of the $\rho, a, b, \eta, \omega, N$ and $\Delta$ 
particles have been considered by taking the mass difference between the 
excited and the ground state particle from PDG and adding this splitting 
to the ground state mass measured on the lattice.
Since this analysis is intended to stay on a qualitative level only, given 
the unknown systematics, we do not consider errors on the hadron masses 
either taken from PDG or from a lattice study. Masses for other particles 
than listed in above table have been set to their PDG values.

\begin{figure}[htb] 
\centering
\includegraphics[height=0.34\textwidth]{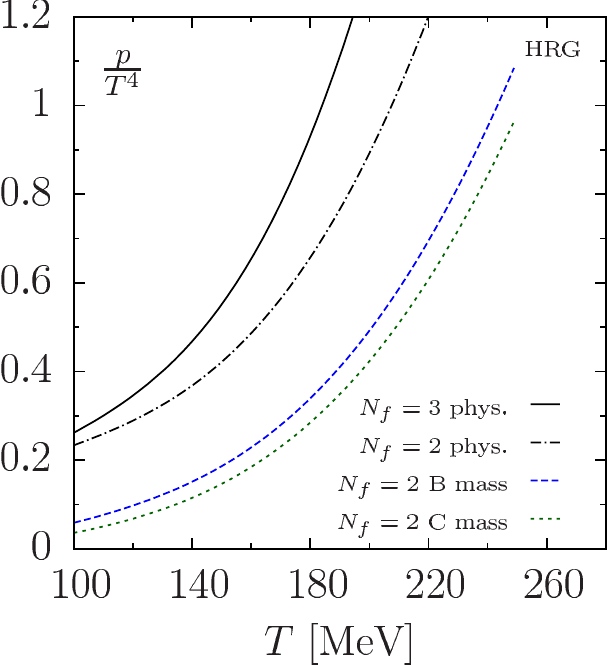} \hspace{0.1\textwidth}
\includegraphics[height=0.34\textwidth]{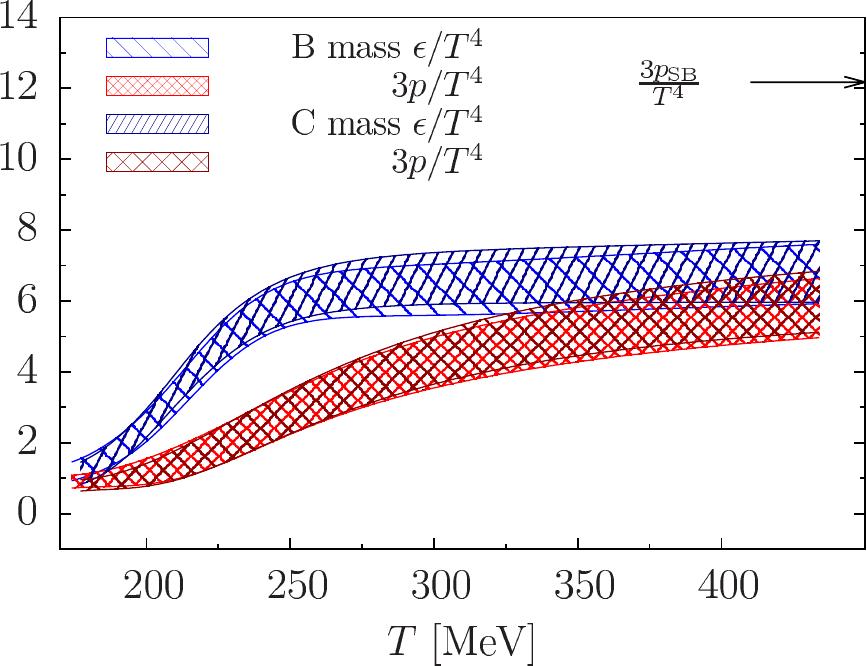} \hfill
\caption{\links The pressure as obtained from several HRG models is shown. 
See text for details. \rechts The pressure and energy density for the B mass
($m_\pi \sim 360$ MeV) and for the C mass ($m_\pi \sim 430$ MeV) 
as obtained when using the lattice HRG model pressure for estimating the 
integration constant $p_0$.
}
\label{fig_p_hrg}
\end{figure}

As can be seen from the curve labelled with ``$N_f=2$ HRG lattice`` in 
\Fig{fig_em3p_hrg}, this $N_f=2$ HRG model incorporating the unphysically 
high masses is compatible with our determination of the trace anomaly at
temperatures in the vicinity to the transition both for the B and the C mass.
In the left panel of \Fig{fig_p_hrg} we show curves for the pressure at 
low temperature obtained from the various adaptations of the HRG model 
under consideration. The curves corresponding to the B and C mass we can 
use to fix the value for the integration constant $p_0$. 
At $T=174~\mathrm{MeV}$ we obtain for the B mass $p_0^B=0.302$ and at 
$T=177~\mathrm{MeV}$ we obtain for the C mass $p_0^C=0.267$.
Using these values to start the pressure integral and assuming a 
conservative 20 \% error on $p_0$ in both cases we obtain for the 
integrated pressure and the energy density the curves depicted in the 
right panel of \Fig{fig_p_hrg}.

\section{Comparison with other results} \label{sec:comparison}

\begin{figure}[htb] 
\centering
\includegraphics[height=0.34\textwidth]{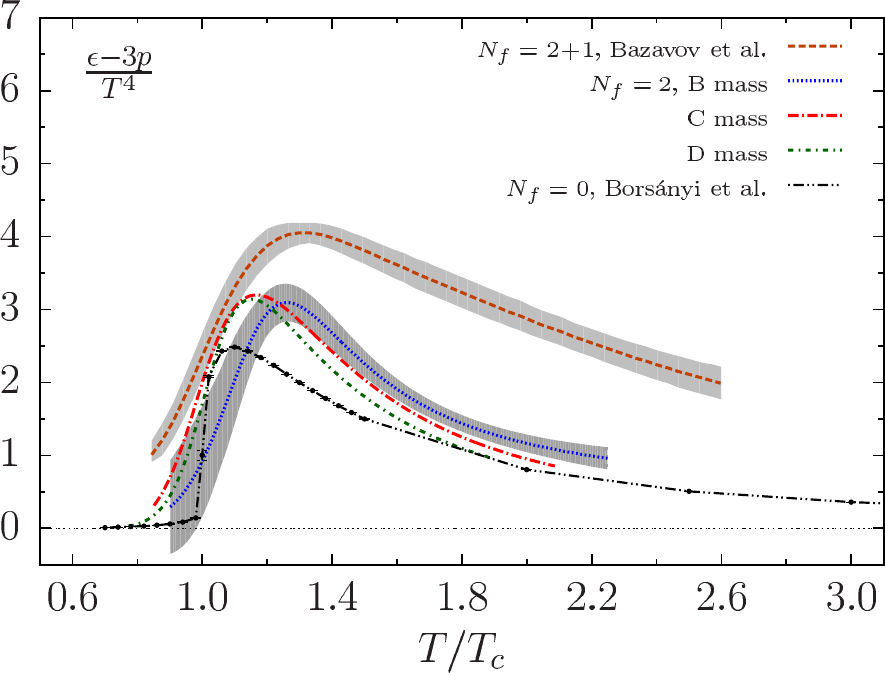}
\caption{A comparison of $I/T^4$ versus $T/T_c$ between 
$N_f=0$ obtained in \Cite{Borsanyi:2012ve}, our data at $N_f=2$ 
for the B mass and $N_f=2+1$ obtained in \Cite{Bazavov:2014pvz}. 
We also show our curves for the C and D ($m_\pi \sim 640$ MeV) masses, 
however, suppressing the errors for better visibility. 
For $T_c$ we use our $T_\chi$ estimates obtained at the largest $N_\tau$ 
available.
}
\label{fig_final}
\end{figure}

Since the trace anomaly is the starting point for all bulk thermodynamics
observables it is natural to choose this quantity for a comparison with 
other results. In our $N_f=2$ study the maximum of the trace anomaly 
has a height of $\sim 3$. Continuum extrapolated results for $N_f = 2+1$ 
at the physical point are reported for stout staggered quarks in 
\Cite{Borsanyi:2010cj}) and for HISQ staggered quarks in 
\Cite{Bazavov:2014pvz}. Both report the maximum of the trace anomaly at height 
$\sim 4$. A study using Wilson quarks together with the fixed scale approach
reports the maximum at a value of $\sim 7.5$ \cite{Umeda:2012er}.
We compare our result for the trace anomaly at the smallest mass with a 
peak height of $\sim 3$ with the data of \Cite{Bazavov:2014pvz} in the 
right panel of \Fig{fig_final}. The data is shown as a function of 
the ratio $T/T_c$, where we use our estimates from Table \ref{tab:tpc} 
at the largest available $N_\tau$ for $T_c$.  
It is also worthwile to compare with the $N_f=0$ case 
for which the EoS was computed in \Cite{Boyd:1996bx} and more recently with 
increased precision in \Cite{Borsanyi:2012ve}. The continuum extrapolated 
data taken from Table 1 of latter reference is also shown in  
\Fig{fig_final}. We have connected data points with lines to 
guide the eye. The $N_f=0$ peak value is smaller than for $N_f=2$, 
and the falling edge of the trace anomaly stays below our interpolation 
for the B mass. The curves from our two larger quark masses seem to 
approach the $N_f=0$ curve at large temperature.

When including a non-zero $p_0$ as a starting point for the pressure 
integration we obtain at  $T = 2~T_c$ a value of $p/p_\mathrm{SB} = 0.45(7)$ 
and $0.48(7)$  for the B and C mass, respectively. 
These values are slightly smaller than 
those computed with $N_f=2+1$ at the physical point in References 
\cite{Borsanyi:2010cj} and \cite{Bazavov:2014pvz}.

\section{Conclusions} \label{sec:conclusions}

In this work we have presented a calculation of QCD thermodynamics 
with two degenerate flavors of Wilson twisted mass quarks. To our knowledge 
this is the first work at $N_f = 2$ providing a continuum limit estimate of
thermodynamics. Moreover, our work constitutes the first fully systematic
determination of the trace anomaly using a Wilson type quark discretization.
Since we were not (yet) able to work at the physical value of the pion mass, 
we have conducted the calculation at three values of unphysically large 
pion mass. Comparing the results as a function of $T/T_c$ we found only 
little residual mass dependence for the trace anomaly, while $T_c$ itself 
decreases with smaller mass as was seen in \Cite{Burger:2011zc}.
Here we investigated the pseudo-critical temperature further at several 
lattice spacings for each considered (charged) pion mass and found no 
significant $N_\tau$ dependence. 

The trace anomaly depends on the quark mass. For a small mass 
interval we find that this dependence is mostly due to a shift in $T_c$, 
since the results expressed as a function of $T/T_c$ show little mass 
dependence. However, there is a clear sensitivity to the matter content 
which is seen by comparing with the quenched determination and results 
with a dynamical strange mass, and lighter quarks. The peak height of the
interaction measure is steadily increasing when enlarging  the number of 
active quark flavors. This suggests that the QGP is more strongly 
interacting when adding more fermions which confirms and extends the 
analysis of \Cite{Liao:2012tw} and \Cite{Miura:2012zqa}.

On the basis of a continuum estimate of the trace anomaly using an 
interpolation ansatz we have calculated by means of the integral method 
(up to an integration constant) the pressure, the energy density and the 
speed of sound in the transition region and up to $\sim (2.0 - 2.5)~T_c$.

We have compared our findings for the trace anomaly at the two lower quark 
mass values in the region of the transition with adaptations of the hadron 
resonance gas model. In order to reproduce (within errors) in these two 
cases of lower quark mass the rising part of the trace anomaly, not only 
hadrons with strangeness had to be disregarded in the model, but also 
the masses of their ground and also excited states had to be 
adapted to match the non-physical masses used in our simulations. 
With these adaptations to the HRG model we found it to agree with our 
trace anomaly results. Given this agreement we have extracted from the 
HRG model a value of the integration constant for the pressure at our 
smallest available temperatures.

\begin{acknowledgments}
F.B.~ and M.M.P.~acknowledge support by the Corroborative Research Center 
SFB / Transregio 9 {\it Computational Particle Physics}. We are grateful 
to the HLRN supercomputing centers Berlin and Hannover for generous 
allocations of computation time. We thank M.~Kirchner, C.~Pinke, C.~Urbach 
and L.~Zeidlewicz for simulating part of the B12, C12 and B10 gauge field
ensembles used in this analysis as well as O.~Philipsen for fruitful  
discussions and hints. We are grateful to the European Twisted Mass 
Collaboration and its members for generating and providing $T=0$ gauge 
field ensembles and for continuous support. We thank M.~Wagner for 
providing us with the code for the determination of $\rnulla$ as well as 
E.~Garcia-Ramos for providing us with information about the $\beta=4.35, 
a \mu = 0.00175$ ETMC gauge ensemble prior to publication. 
For the generation of gauge field configurations we have used 
tmLQCD \cite{Jansen:2009xp,Abdel-Rehim:2013wba} on massively-parallel 
systems as well as on GPU-equipped PC-clusters. Statistical analyses were 
done with the help of R\cite{Rpackage}.
\end{acknowledgments}

\begin{appendix}
\section{Symanzik expansion of $\vevsub{\bar \chi \chi}$}  
\label{sec_m0symanzik}

Starting from \Eq{eq:traceanomaly} we need to measure a term 
$\vevsub{\sum_x \bar \chi_x \chi_x}$ stemming from the $m$-derivative 
of the action \Eq{eq:tmaction}. Since the sum is composed of short 
distance contributions and the operator may mix with different operators 
of the same symmetry the usual arguments of automatic $\mathcal{O}(a)$ 
improvement does not necessarily hold. In \cite{Burger:2014ada} the 
Symanzik expansion of the vacuum polarization tensor has been studied. 
We rely for the following argument on the characterization of operators 
in terms of symmetry transformations achieved there.

At maximal twist the symmetry transformations of $\vevsub{\bar \chi \chi}$ 
(we suppress the sum over spacetime in what follows) read:
\begin{center}
{\small
\begin{tabular}{c|c|c|c|c|c|c|c}
& $\mathcal{P}_{1/2}$ & $\mathcal{P}_{[\mu \rightarrow -\mu]}$ 
& $\mathcal{T}_{1/2}$ & $\mathcal{T}_{[\mu \rightarrow -\mu]}$ 
& $\mathcal{C}$ & $\mathcal{P} \mathcal{D} [-m][-r]$ 
& $\ronetwofive \mathcal{D} [-\mu]$\\
 \hline
  $\bar \chi \chi$ & +1 & +1 & +1 & +1 & +1 & -1 & +1\\
\end{tabular} \;.
}
\end{center}
Being an operator of mass dimension three $\bar \chi \chi$  will mix 
with operators of mass dimension lower or equal to three under 
renormalization which have the same symmetry transformation properties.
The local operators having the same symmetry properties as 
$\bar \chi \chi$  up to dimension 4 read:
\begin{equation}
 \left \{ r , m_q, r m_q^2, r m_q \bar \chi \chi, 
          r \mu_q \bar \chi \gamma_5 \tau^3 \chi \right \} \,,
\end{equation}
where $m_q$ is to be considered as the subtracted quark mass 
$m_q = m - m_c$ with $m_c$ denoting the critical value.

Using the above set a finite subtracted operator can be constructed as:
\begin{equation}
\begin{split}
 \bar \chi \chi_{R} &= Z^{\bar \chi \chi} \bar \chi \chi + 
   r \frac{Z^r}{a^3} + m_q \frac{Z^{m_q}}{a^2} + m_q^2 \frac{Z^{m_q^2}}{a} 
   + a r m_q Z^{rm_q} + a r \mu_q Z^{r\mu_q^2} + \mathcal{O}(a^2) \,.
\end{split}
\end{equation}
The expansion becomes complete once also the effective Lagrangian is 
expanded:
\begin{equation}
 \mathcal{L}_{\mathrm{eff}} = \mathcal{L}_{4} + a \mathcal{L}_{5} +  
          a^2 \mathcal{L}_{6} + a^3 \mathcal{L}_{7} +\mathcal{O}(a^4) \,.
\end{equation}
From the expansion of the effective action $S$ in the Boltzmann weight
\begin{equation}
\begin{split}
 \exp{(-S)} &= \exp{(-S_4)} {\Big \{ } 1 - a S_5 + 
               a^2 \left(\frac{1}{2} S_5^2 - S_6 \right ) \\
            &\hphantom{\exp{(S_4)} {\Big \{ }} + 
                a^3 \left (-\frac{1}{6} S_5^3 + S_5 S_6 - S_7 \right ) 
                + \mathcal{O}(a^4)  {\Big\} } \\
\end{split}
\end{equation}
further terms arise in the Symanzik expansion of the operator, 
where the notation $S_i = \int \mathcal{L}_i(x) d^4x $ has been used.
Keeping only terms up to order $\mathcal{O}(a)$ and restricting to the 
case $m_q \equiv 0$ that is implied by the maximal twist 
condition the combined Symanzik expansion is given by:
\begin{equation}
\begin{split}
 \vev{\bar \chi \chi}_{\mathrm{R}} &= 
  Z^{\bar \chi \chi} \vev{\bar \chi \chi}_0 + r \frac{Z^r}{a^3} \vev{1}_0 
   + r \frac{Z^r}{a^2} \vev{-\mathcal{S}_{5}}_0 \\
  &+ r \frac{Z^r}{a} \left ( \vev{-\mathcal{S}_{6} + 
     \frac{1}{2} S_5^2}_0 \right ) \\
  &+ r Z^r \left ( \vev{\mathcal-{S}_{7} + S_5 S_6 - 
     \frac{1}{6} S_5^3}_0 \right ) + \mathcal{O}(a) \,.
\end{split}
\end{equation}
The divergent terms $\sim a^{-n}$ need to be cared for by subtracting 
the $T=0$ result. The two finite terms $\vev{\bar \chi \chi}_0$ and 
$r Z^r \vev{\mathcal{L}_{7}}_0$ vanish due to $\ronetwofive$ symmetry 
of $\mathcal{L}_{4}$ over which the terms are averaged, since they are 
$\ronetwofive$-odd. After $T=0$ subtraction we thus note that the 
remaining terms are lattice artefacts of $\mathcal{O}(a)$ and higher. 

For the trace anomaly we however need to consider the bare dimensionless 
lattice operator $a^3 \vev{\bar \chi \chi}_0$ instead of the fully 
subtracted and multiplicatively renormalized operator discussed above. 
Restricting the expansion to $\mathcal{O}(a^2)$ precision we obtain 
\begin{equation}
\begin{split}
 \frac{a^3}{Z^{\bar \chi \chi}} \vev{\bar \chi \chi}_{sub} &= 
 a^3 \vev{\bar \chi \chi}_0 + r \tilde Z^r \vev{1}_0 + 
 a r \tilde Z^r \vev{-\mathcal{S}_{5}}_0 \\
 &+ r a^2 \tilde Z^r \left ( \vev{-\mathcal{S}_{6} + 
    \frac{1}{2} S_5^2}_0 \right ) + \mathcal{O}(a^3) \,,
\end{split}
\end{equation}
where we have absorbed the $Z^{\bar \chi \chi}$ on the right hand side 
into the definition of the $Z$-factors of the operators. Since the bare 
operator $\bar \chi \chi$ is an $\ronetwofive$-odd (or equally a 
twisted parity odd)  operator its expectation value with respect 
to the $\ronetwofive$-symmetric (twisted parity even) continuum twisted 
mass action vanishes. The constant piece proportional to $\vev{1}_0$ 
will be eliminated by the $T=0$ subtraction of the trace anomaly. 
The remaining terms are lattice artefacts that can be further 
restricted to being of order $\mathcal{O}(a^2)$ and higher upon 
noting that also the contributions to $S_5$ are $\ronetwofive$-odd. 
The $T=0$ subtracted contribution to the trace anomaly from the $m$ 
derivative of the action is thus seen to be vanishing in the continuum 
limit at $\mathcal{O}(a^2)$ and can (and also should) be disregarded 
in the evaluation right from the beginning.

\newpage
\section{Tables} \label{sec:simtables}

\vspace*{2cm}
\begin{table*}[hb]
\footnotesize
 \begin{center}
    \begin{tabular}{c|c|c|c|c|c|c|c|c}
$N_\tau \ N_\sigma$ & $\beta$ &  $\kappa$  & $a\mu$ & $a m_\mathrm{PS}$ & $m_\mathrm{PS} L$ &  $\vev{S_g}$ & $\vev{S_f}~(\cdot10^{2})$ & $\rnulla$   \\
       \hline\hline
$48 \ 24$& 3.80 & 0.164111 & 0.00600 & 0.1852(9) & 4.4 & 5.34639(94) & $ 4.467(16)$ & 4.321(32) \\
$48 \ 24$& 3.80 & 0.164111 & 0.00800 & 0.2085(8) & 5.0 & 5.34747(60) & $ 5.590(09)$ & 4.440(34) \\
$48 \ 24$& 3.80 & 0.164111 & 0.01100 & 0.2424(5) & 5.8 & 5.34900(82) & $ 7.164(07)$ & 4.362(21) \\
$48 \ 24$& 3.80 & 0.164111 & 0.01650 & 0.2957(5) & 7.1 & 5.34887(22) & $10.001(08)$ & 4.264(14) \\
$64 \ 32$& 3.90 & 0.160856 & 0.00300 & 0.1167(4) & 3.7 & -           & -            & -\\
$64 \ 32$& 3.90 & 0.160856 & 0.00400 & 0.1338(2) & 4.3 & 5.47032(15) & $ 2.753(07)$ & -\\
$48 \ 24$& 3.90 & 0.160856 & 0.00400 & -         & -   & 5.47012(33) & $ 2.737(08)$ & 5.196(28) \\
$32 \ 16$& 3.90 & 0.160856 & 0.00400 & -         & -   & 5.47108(54) & $ 2.490(21)$ & -\\
$48 \ 24$& 3.90 & 0.160856 & 0.00640 & 0.1694(4) & 4.1 & 5.47027(17) & $ 3.923(09)$ & 5.216(27) \\
$48 \ 24$& 3.90 & 0.160856 & 0.00850 & 0.1940(5) & 4.7 & 5.47011(27) & $ 4.926(10)$ & 5.130(28) \\
$48 \ 24$& 3.90 & 0.160856 & 0.01000 & 0.2100(5) & 5.0 & 5.47074(18) & $ 5.654(11)$ & 5.143(25) \\
$48 \ 24$& 3.90 & 0.160856 & 0.01500 & 0.2586(7) & 6.2 & 5.47003(27) & $ 8.030(10)$ & 5.039(21) \\
$64 \ 32$& 4.05 & 0.157010 & 0.00300 & 0.1038(6) & 3.3 & 5.63404(19) & $1.691(05)$  & 6.584(34) \\
$64 \ 32$& 4.05 & 0.157010 & 0.00600 & 0.1432(6) & 4.6 & 5.63419(08) & $3.019(03)$  & 6.509(38) \\
$64 \ 32$& 4.05 & 0.157010 & 0.00800 & 0.1651(5) & 5.3 & 5.63419(14) & $3.879(05)$  & 6.494(36) \\
$64 \ 32$& 4.05 & 0.157010 & 0.01200 & 0.2025(8) & 6.5 & 5.63400(06) & $5.612(04)$  & 6.284(22) \\
$96 \ 48$& 4.20 & 0.154073 & 0.00200 & 0.0740(3) & 3.6 & 5.78133(06) & $1.0013(29)$ & 8.295(45) \\
$64 \ 32$& 4.20 & 0.154073 & 0.00650 & 0.1326(5) & 4.2 & 5.78130(05) & $2.8200(33)$ & 8.008(29) \\
$64 \ 32$& 4.35 & 0.151740 & 0.00175 & 0.0748(17)& 2.4 & 5.91477(04) & $0.7363(48)$ & 9.9(2) \\
     \end{tabular}
 \end{center}
\caption{Parameters and results for $T=0$ ETMC generated gauge ensembles 
used in this analysis. For further details we refer the reader 
to \Cites{Baron:2009wt,Jansen:2011vv}. We show results for the pseudoscalar 
mass, $\rnulla$, the gauge action $\vev{S_g}$ and the fermion action $\vev{S_f}$. 
The latter two have been evaluated fully using the available statistics.}
\label{tab_T0_ETM}
\end{table*}
\begin{table*}
\footnotesize
 \begin{center}
    \begin{tabular}{c|c|c|c|c|c|c|c|c|c|c}
$N_\tau \ N_\sigma$ & $\beta$ &  $\kappa$  & $a\mu$ & $a m_\mathrm{PS}$ & $m_\mathrm{PS} L$ & $a m_{\mathrm{PCAC}}$ &  $\vev{S_g}$& $\vev{S_f}~(\cdot10^{2})$ & $\rnulla$ & TU  \\
       \hline\hline
$24 \ 16$& 3.65 & 0.170250   & 0.01200  & 0.302(6) & 4.8 & $7(9)\cdot10^{-4}$     & -           & -            & 3.437(33) & 17396\\
$40 \ 16$& 3.65 & 0.170200   & 0.02517  & 0.425(7) & 6.8 & $2(2)\cdot10^{-3}$     & -           & -            &  3.249(64) & 3610 \\
$24 \ 16$& 3.70 & 0.168062   & 0.00900  & 0.254(8) & 4.0 & $-3.6(1.0)\cdot10^{-3}$& 5.2231(12)  & $ 7.356(29)$ &  3.853(59) & 29042\\
$24 \ 16$& 3.70 & 0.168062   & 0.01055  & 0.268(6) & 4.3 & $-2.6(7)\cdot10^{-3}$  & 5.2212(13)  & $ 8.331(23)$ &  3.706(116)& 15126\\
$40 \ 20$& 3.70 & 0.168092   & 0.02406  & 0.397(7) & 7.9 & $-4.8(9)\cdot10^{-3}$  & 5.21675(32) & $16.085(16)$ &  3.673(40) & 7359 \\
$20 \ 20$& 3.72 & 0.167216   & 0.00724  &   -      & -   & -                      & -           & -            &  3.895(51) & 5637 \\
$20 \ 20$& 3.72 & 0.167229   & 0.02342  &   -      & -   & -                      & 5.24341(52) & $15.236(22)$ &  3.721(38) & 3017 \\
$20 \ 20$& 3.74 & 0.166401   & 0.02279  &   -      & -   & -                      & 5.26912(46) & $14.396(15)$   &  -         & 6237 \\
$24 \ 16$& 3.76 & 0.165607   & 0.00689  & 0.208(5) & 3.3 & $ 5(6)\cdot10^{-4}$    & 5.29840(53) & $ 5.291(19)$ &  4.186(52) & 52665\\
$24 \ 16$& 3.76 & 0.165607   & 0.00979  & 0.246(6) & 3.9 & $-1.1(7)\cdot10^{-3}$  & 5.29826(54) & $ 6.979(10)$ &  4.058(97) & 20000\\
$40 \ 20$& 3.76 & 0.165608   & 0.02218  & 0.354(9) & 7.0 & $-2.1(9)\cdot10^{-3}$  & 5.29581(24) & $13.652(08)$ &  3.972(24) & 7405 \\
$20 \ 20$& 3.78 & 0.164844   & 0.02158  &   -      & -   & -                      & 5.32171(49) & $12.916(15)$ &  -         & 2900 \\
$32 \ 20$& 3.80 & 0.164111   & 0.00655  &   -      &     & -                      & 5.34846(27) & $ 4.787(08)$ &  -         & 24343 \\
$40 \ 20$& 3.80 & 0.164111   & 0.02100  & 0.338(4) & 6.7 & $-1.8(9)\cdot10^{-3}$  & 5.34767(28) & $12.274(11)$ &  4.166(51) & 5319 \\
$48 \ 24$& 3.82 & 0.163407   & 0.00639  &   -      & -   & -                      & 5.37469(76) & $ 4.538(12)$ &  -         & 2263 \\
$20 \ 20$& 3.82 & 0.163407   & 0.02043  &   -      & -   & -                      & 5.37207(49) & $11.672(13)$ &  -         & 3030 \\
$48 \ 24$& 3.84 & 0.162731   & 0.00623  &   -      & -   & -                      & 5.39978(55) & $ 4.281(10)$ &  -         & 1941 \\
$20 \ 20$& 3.84 & 0.162731   & 0.01989  &   -      & -   & -                      & 5.39851(37) & $11.066(19)$ &  -         & 2541 \\
$48 \ 24$& 3.85 & 0.162403   & 0.00600  & 0.175(2) & 4.2 & $-1(4)\cdot10^{-4}$    & 5.41145(75) & $ 4.076(09)$ &  4.711(53) & 1244  \\
$48 \ 24$& 3.85 & 0.162403   & 0.00893  & 0.208(4) & 5.0 & $0.2(4.4)\cdot10^{-4}$ & 5.41094(17) & $ 5.573(06)$ &  4.684(43) & 4881 \\
$40 \ 20$& 3.85 & 0.162403   & 0.01962  & 0.311(6) & 6.2 & $-1.9(8)\cdot10^{-3}$  & 5.41025(15) & $10.799(11)$ &  4.550(52) & 4880 \\
$48 \ 24$& 3.86 & 0.162081   & 0.00617  & 0.174(2) & 4.1 & $-3(3)\cdot10^{-4}$    & 5.42323(21) & $ 4.092(06)$ &  -         & 10054\\
$20 \ 20$& 3.86 & 0.162081   & 0.01935  & -        & -   & -                      & 5.42195(29) & $10.553(14)$ &  -         & 3003 \\
$20 \ 20$& 3.87 & 0.161766   & 0.01909  & -        & -   & -                      & 5.43396(20) & $10.285(13)$ &  -         & 4517 \\
$48 \ 24$& 3.88 & 0.161457   & 0.00600  & 0.168(5) & 4.0 & $-7(4)\cdot10^{-4}$    & 5.44729(19) & $ 3.857(04)$ &  -         & 9528 \\
$20 \ 20$& 3.88 & 0.161457   & 0.01883  & -        & -   & -                      & 5.44576(24) & $10.060(19)$ &  -         & 4032 \\
$40 \ 20$& 3.90 & 0.160856   & 0.01833  & 0.292(6) & 5.8 & $-1.9(9)\cdot10^{-3}$  & 5.46948(18) & $ 9.589(09)$ & 4.842(44)  & 2522 \\
$20 \ 20$& 3.92 & 0.160278   & 0.01784  & -        & -   & -                      & 5.49305(23) & $ 9.111(12)$ &  -         & 2899 \\
$48 \ 24$& 3.93 & 0.159998   & 0.00561  & 0.158(3) & 3.7 & $-1(3)\cdot10^{-4}$    & 5.50477(22) & $ 3.345(04)$ & 5.447(61)  & 9324 \\
$48 \ 24$& 3.93 & 0.159998   & 0.00801  & 0.182(8) & 4.4 & $-8(7)\cdot10^{-3}$    & 5.50495(24) & $ 4.486(08)$ & 5.367(72)  & 2817 \\
$20 \ 20$& 3.93 & 0.159997   & 0.01759  &   -      & -   & -                      & 5.50434(18) & $ 8.917(25)$ & 5.324(83)  & 4437 \\
$20 \ 20$& 3.94 & 0.159722   & 0.01736  &   -      & -   & -                      & 5.51565(25) & $ 8.717(13)$ &  -         & 2809 \\
$48 \ 24$& 3.95 & 0.159452   & 0.00546  & 0.151(3) & 3.6 & $2(2)\cdot10^{-4}$     & 5.52714(11) & $ 3.185(05)$ &  -         & 6955 \\
$20 \ 20$& 3.96 & 0.159187   & 0.01689  & -        & -   & -                      & 5.53763(21) & $ 8.330(14)$ &  -         & 2045 \\
$64 \ 32$& 3.97 & 0.158927   & 0.00531  & 0.144(1) & 4.6 & $-5(2)\cdot10^{-4}$    & 5.54940(10) & $ 3.013(06)$ & 5.809(112) & 2299 \\
$48 \ 24$& 3.97 & 0.158926   & 0.00752  & 0.176(7) & 4.2 & $-6(7)\cdot10^{-4}$    & 5.54901(10) & $ 4.036(04)$ & 5.733(59)  & 4800 \\
$40 \ 20$& 3.97 & 0.158926   & 0.01666  & 0.263(4) & 5.2 & $-1.7(8)\cdot10^{-3}$  & 5.54880(12) & $ 8.151(10)$ & 5.455(55)  & 4211 \\
$20 \ 20$& 3.98 & 0.158671   & 0.01644  & -        & -   & -                      & 5.55936(21) & $ 7.972(10)$ &  -         & 2806 \\
$48 \ 24$& 3.99 & 0.158421   & 0.00517  & 0.141(4) & 3.3 & $-2(3)\cdot10^{-4}$    & 5.57097(11) & $ 2.856(05)$ &  -         & 5538 \\
$64 \ 32$& 4.01 & 0.157933   & 0.00503  & 0.135(2) & 4.3 & $-3(3)\cdot10^{-4}$    & 5.59268(16) & $ 2.722(10)$ &  -         & 926  \\
$48 \ 24$& 4.01 & 0.157933   & 0.00718  & 0.164(4) & 3.9 & $-3(4)\cdot10^{-4}$    & 5.59207(10) & $ 3.690(05)$ &  -         & 5264 \\
$40 \ 20$& 4.05 & 0.157010   & 0.01520  & 0.233(8) & 4.6 & $-3(2)\cdot10^{-3}$    & 5.63380(09) & $ 6.974(09)$ & 6.233(76)  & 4180 \\
$64 \ 32$& 4.10 & 0.155945   & 0.00445  & 0.117(2) & 3.7 & $2(1)\cdot10^{-4}$     & 5.68485(06) & $ 2.192(05)$ &  -         & 2090 \\
$20 \ 20$& 4.10 & 0.155946   & 0.01431  & -        & -   & -                      & 5.68453(18) & $ 6.334(03)$ &  -         & 1485 \\
$48 \ 24$& 4.20 & 0.154073   & 0.01000  & 0.16(2)  & 3.8 & $4(12)\cdot10^{-4}$    & 5.78135(11) & $ 4.205(08)$ & 7.6(2)     & 810  \\
$40 \ 20$& 4.20 & 0.154073   & 0.01270  & 0.20(2)  & 4.0 & $-3(2)\cdot10^{-3}$    & 5.78114(08) & $ 5.254(08)$ &  -         & 3432 \\
$48 \ 24$& 4.35 & 0.151740   & 0.00600  & 0.14(3)  & 3.3 & $-8(20)\cdot10^{-4}$   & 5.91479(09) & $ 2.363(06)$ & 9.61(49)   & 977  \\
$40 \ 20$& 4.35 & 0.151740   & 0.01050  & 0.176(8) & 3.5 & $-0.2(9)\cdot10^{-3}$  & 5.91485(08) & $ 4.069(18)$ & 10.19(55)  & 3907 \\
     \end{tabular}
 \end{center}
\caption{Simulation parameters for the $T=0$ runs. We show results for the
pseudoscalar and PCAC mass as well as $\rnulla$ where calculated.}
\label{tab_T0}
\end{table*}
\begin{table*}
\footnotesize
 \begin{center}
    \begin{tabular}{c c|c|c|c|c|c|c|c|c}
       $N_\tau$ & $N_\sigma$ & $\beta$ &  $\kappa$  & $a\mu$ & $T [\mathrm{MeV}]$ & $\Real{(L)}$ & $\vev{S_g}$ & $\vev{S_f}~(\cdot10^{2})$ & TU  \\
       \hline\hline
12 & 32 & 3.86 & 0.162081 & 0.00617 &	175 & $5.9(3)\cdot10^{-4}$  &5.42332(22) &  $3.937(08)$& 16697 \\
   &    & 3.88 & 0.161457 & 0.00600 &	181 & $7.3(3)\cdot10^{-4}$  &5.44716(13) &  $3.691(06)$& 17375 \\
   &    & 3.90 & 0.160856 & 0.00584 &	186 & $8.7(2)\cdot10^{-4}$  &5.47070(17) &  $3.440(07)$& 14249 \\
   &    & 3.93 & 0.159998 & 0.00561 &	195 & $1.24(3)\cdot10^{-3}$ &5.50501(13) &  $3.114(11)$& 12099 \\
   &    & 3.95 & 0.159452 & 0.00546 &	201 & $1.51(4)\cdot10^{-3}$ &5.52741(12) &  $2.930(10)$& 7878  \\
   &    & 3.97 & 0.158927 & 0.00531 &	208 & $1.95(3)\cdot10^{-3}$ &5.54952(12) &  $2.724(10)$& 9653 \\
   &    & 3.99 & 0.158421 & 0.00517 &	214 & $2.20(4)\cdot10^{-3}$ &5.57121(10) &  $2.557(13)$& 8968  \\
   &    & 4.01 & 0.157933 & 0.00503 &	220 & $2.70(5)\cdot10^{-3}$ &5.59285(06) &  $2.391(10)$& 15223 \\
   &    & 4.04 & 0.157235 & 0.00689 &	230 & $3.39(5)\cdot10^{-3}$ &5.62441(08) &  $2.186(07)$& 6080 \\
   &    & 4.07 & 0.156573 & 0.00463 &	241 & $4.07(6)\cdot10^{-3}$ &5.65537(10) &  $2.030(07)$& 3359\\
   &    & 4.10 & 0.155945 & 0.00445 &	251 & $4.84(5)\cdot10^{-3}$ &5.68538(06) &  $1.894(04)$& 15073\\
   &    & 4.15 & 0.154969 & 0.00422 &	270 & $6.17(7)\cdot10^{-3}$ &5.73446(07) &  $1.736(03)$& 4080 \\
   &    & 4.20 & 0.154073 & 0.00396 &	290 & $7.57(8)\cdot10^{-3}$ &5.78177(06) &  $1.5828(19)$& 4640 \\
   &    & 4.25 & 0.153247 & 0.00372 &	311 & $9.17(7)\cdot10^{-3}$ &5.82769(07) &  $1.4510(23)$& 4160 \\
   &    & 4.35 & 0.151740 & 0.00316 &	356 & $1.22(1)\cdot10^{-2}$ &5.91511(05) &  $1.1852(17)$& 4334 \\
\hline
10 & 32 & 3.76 & 0.165607 & 0.00689 &	178 & $1.53(3)\cdot10^{-3}$ &5.29836(42) &  $5.138(08)$& 18438\\
   &    & 3.78 & 0.164844 & 0.00672 &	184 & $1.87(4)\cdot10^{-3}$ &5.32338(63) &  $4.756(11)$& 10385\\
   &    & 3.80 & 0.164111 & 0.00655 &	190 & $2.29(4)\cdot10^{-3}$ &5.34970(26) &  $4.408(11)$& 11692\\
   &    & 3.82 & 0.163407 & 0.00639 &	197 & $2.72(6)\cdot10^{-3}$ &5.37499(29) &  $4.085(15)$& 7811 \\
   &    & 3.84 & 0.162731 & 0.00623 &	203 & $3.34(4)\cdot10^{-3}$ &5.40005(15) &  $3.772(15)$& 9433 \\
   &    & 3.88 & 0.161457 & 0.00600 &	217 & $4.63(5)\cdot10^{-3}$ &5.44832(15) &  $3.233(61)$& 7945 \\
   &    & 3.90 & 0.160856 & 0.00600 &	224 & $5.75(8)\cdot10^{-3}$ &5.47185(13) &  $3.101(20)$& 2987 \\
   &    & 3.93 & 0.159998 & 0.00600 &	234 & $7.17(10)\cdot10^{-3}$&5.50601(24) &  $2.967(14)$& 4025 \\
   &    & 3.95 & 0.159452 & 0.00545 &	242 & $8.03(8)\cdot10^{-3}$ &5.52882(17) &  $2.578(08)$& 1971\\   
   &    & 3.97 & 0.158926 & 0.00529 &	249 & $8.53(9)\cdot10^{-3}$ &5.55047(09) &  $2.469(06)$& 7276 \\
   &    & 4.01 & 0.157933 & 0.00503 &	265 & $1.06(2)\cdot10^{-2}$ & -          &  -          & 2720 \\
   &    & 4.05 & 0.157010 & 0.00478 &	281 & $1.25(1)\cdot10^{-2}$ &5.63524(06) &  $2.060(03)$& 8716 \\   
   &    & 4.10 & 0.155945 & 0.00449 &	302 & $1.45(2)\cdot10^{-2}$ &5.68587(10) &  $1.873(04)$& 2211 \\
   &    & 4.20 & 0.154073 & 0.00396 &	348 & $2.02(2)\cdot10^{-2}$ &5.78199(07) & $1.5640(05)$& 4000 \\
   &    & 4.35 & 0.151740 & 0.00326 &	428 & $2.86(2)\cdot10^{-2}$ &5.91510(09) & $1.2199(04)$& 2235 \\
\hline
8 & 28 & 3.76 & 0.165607 & 0.00689 &	222 & $1.21(2)\cdot10^{-2}$& 5.30280(42) &  $4.204(14)$& 4350 \\ 
  &    & 3.80 & 0.164111 & 0.00655 &	238 & $1.58(2)\cdot10^{-2}$& 5.35447(28) &  $3.584(10)$& 4500 \\ 
  &    & 3.85 & 0.162401 & 0.00615 &	258 & $2.01(2)\cdot10^{-2}$& 5.41578(15) &  $3.082(07)$& 4444 \\ 
  &    & 3.90 & 0.160856 & 0.00578 &	280 & $2.46(2)\cdot10^{-2}$& 5.47404(21) &  $2.722(04)$& 3007 \\ 
  &    & 3.97 & 0.158934 & 0.00529 &	311 & $3.08(2)\cdot10^{-2}$& 5.55181(13) &  $2.348(03)$& 3148 \\ 
  &    & 4.01 & 0.157955 & 0.00503 &	331 & $3.45(2)\cdot10^{-2}$& 5.59437(12) & $2.1692(20)$& 2746 \\ 
  &    & 4.05 & 0.157010 & 0.00479 &	351 & $3.80(2)\cdot10^{-2}$& 5.63594(11) & $2.0119(07)$& 3792 \\ 
  &    & 4.10 & 0.155952 & 0.00449 &	377 & $4.27(2)\cdot10^{-2}$& 5.68652(10) & $1.8372(04)$& 3581 \\ 
  &    & 4.20 & 0.154073 & 0.00396 &	435 & $5.12(2)\cdot10^{-2}$& 5.78247(11) & $1.5462(03)$& 3750 \\ 
  &    & 4.35 & 0.151740 & 0.00328 &	535 & $6.56(2)\cdot10^{-2}$& 5.91527(08) & $1.1681(02)$& 4200 \\
\hline 
6 & 32 & 3.80 & 0.164111 & 0.00655 &	317 & $7.17(3)\cdot10^{-2}$ & 5.36235(17) &  $3.2125(10)$& 2926\\ 
  &    & 3.82 & 0.163406 & 0.00639 &	328 & $7.38(3)\cdot10^{-2}$ & 5.38576(20) &  $3.0758(22)$& 2085\\ 
  &    & 3.84 & 0.162730 & 0.00623 &	339 & $7.70(2)\cdot10^{-2}$ & 5.40928(22) &  $2.9446(07)$& 1578\\ 
  &    & 3.86 & 0.162080 & 0.00608 &	350 & $7.98(2)\cdot10^{-2}$ & 5.43233(21) &  $2.8243(09)$& 1611\\ 
  &    & 3.90 & 0.160856 & 0.00578 &	373 & $8.60(2)\cdot10^{-2}$ & 5.47760(16) &  $2.6033(05)$& 2337\\ 
  &    & 3.97 & 0.158934 & 0.00529 &	415 & $9.62(2)\cdot10^{-2}$ & 5.55432(15) &  $2.2734(03)$& 2034\\ 
  &    & 4.05 & 0.157010 & 0.00479 &	468 & $1.076(2)\cdot10^{-1}$& 5.63805(16) &  $1.9634(02)$& 2151\\ 
  &    & 4.10 & 0.155952 & 0.00449 &	503 & $1.15(3)\cdot10^{-1}$ & 5.68829(20) &  $1.7979(02)$& 1025\\
  &    & 4.20 & 0.154073 & 0.00396 &	580 & $1.284(1)\cdot10^{-1}$& 5.78360(06) &  $1.5174(01)$& 7440\\ 
\hline
4 & 32 & 3.80 & 0.164111 & 0.00655 &	476 & $2.507(1)\cdot10^{-1}$& 5.38475(23) &  $2.9518(03)$& 2528\\ 
  &    & 3.86 & 0.162080 & 0.00608 &	525 & $2.605(1)\cdot10^{-1}$& 5.45140(20) &  $2.6245(02)$& 2233\\ 
  &    & 3.90 & 0.160856 & 0.00578 &	559 & $2.669(1)\cdot10^{-1}$& 5.49520(32) &  $2.4324(02)$& 1737\\ 
  &    & 3.97 & 0.158934 & 0.00529 &	623 & $2.782(1)\cdot10^{-1}$& 5.56926(18) &  $2.1392(02)$& 1547\\ 
    \end{tabular}
 \end{center}
\caption{Simulation parameters for the B mass ensembles. Results for the 
bare Polyakov loop, the gauge action $\vev{S_g}$ and the fermion action 
abbreviated as $\vev{S_f}$. TU denotes the number of Monte Carlo time units
simulated.}
\label{tab_Bensemble}
\end{table*}
\begin{table*}
\footnotesize
 \begin{center}
    \begin{tabular}{c c|c|c|c|c|c|c|c|c}
       $N_\tau$ & $N_\sigma$ & $\beta$ &  $\kappa$  & $a\mu$  & $T [\mathrm{MeV}]$ & $\Real{(L)}$ & $\vev{S_g}$& $\vev{S_f}~(\cdot10^{2})$ & TU  \\
       \hline\hline
12 & 32 & 3.90 & 0.160856 & 0.00821 & 186 & $8.4(5)\cdot10^{-4}$  & 5.47062(16) & $4.654(08)$& 5879\\ 
      & & 3.93 & 0.159997 & 0.00801 & 195 & $1.16(4)\cdot10^{-3}$ & 5.50490(17) & $4.309(13)$& 5180\\ 
      & & 3.95 & 0.159452 & 0.00779 & 201 & $1.35(3)\cdot10^{-3}$ & 5.52722(20) & $4.066(21)$& 5822\\ 
      & & 3.97 & 0.158926 & 0.00752 & 208 & $1.63(3)\cdot10^{-3}$ & 5.54938(13) & $3.818(08)$& 9179\\ 
      & & 3.99 & 0.158421 & 0.00738 & 214 & $2.13(5)\cdot10^{-3}$ & 5.57143(10) & $3.603(12)$& 5151\\ 
      & & 4.01 & 0.157933 & 0.00718 & 220 & $2.48(5)\cdot10^{-3}$ & 5.59288(10) & $3.411(16)$& 3270\\ 
      & & 4.03 & 0.157463 & 0.00699 & 227 & $2.92(7)\cdot10^{-3}$ & 5.61370(09) & $3.241(25)$& 6428\\     
      & & 4.05 & 0.157010 & 0.00680 & 234 & $3.57(8)\cdot10^{-3}$ & 5.63471(09) & $3.063(13)$& 2620\\ 
      & & 4.07 & 0.156573 & 0.00662 & 241 & $4.19(9)\cdot10^{-3}$ & 5.65510(07) & $2.905(07)$& 3916\\ 
      & & 4.10 & 0.155946 & 0.00639 & 251 & $4.92(7)\cdot10^{-3}$ & 5.68525(09) & $2.729(08)$& 2613\\ 
      & & 4.15 & 0.154975 & 0.00599 & 270 & $6.2(1)\cdot10^{-3}$  & 5.73446(07) & $2.464(04)$& 2653\\     
      & & 4.20 & 0.154073 & 0.00563 & 290 & $7.4(2)\cdot10^{-3}$  & 5.78167(08) & $2.2465(17)$& 2627\\
      & & 4.25 & 0.153238 & 0.00528 & 310 & $8.6(2)\cdot10^{-3}$  & 5.82742(07) & $2.0612(22)$& 2807\\
      & & 4.35 & 0.151740 & 0.00466 & 356 & $1.22(2)\cdot10^{-2}$ & 5.91509(07) & $1.7446(10)$& 2718\\      
\hline
10 & 32 & 3.76 & 0.165607 & 0.00979 & 178 & $1.29(3)\cdot10^{-3}$ & 5.29785(36) & $6.846(07)$& 10357\\
   &    & 3.80 & 0.164111 & 0.00956 & 190 & $1.94(3)\cdot10^{-3}$ & 5.34910(26) & $6.142(09)$& 9002\\
   &    & 3.85 & 0.162403 & 0.00893 & 207 & $3.20(6)\cdot10^{-3}$ & 5.41170(20) & $5.170(14)$& 7679\\
   &    & 3.90 & 0.160856 & 0.00821 & 224 & $5.22(8)\cdot10^{-3}$ & 5.47155(11) & $4.267(16)$ & 10065\\  
   &    & 3.93 & 0.159998 & 0.00801 & 234 & $6.76(13)\cdot10^{-3}$& 5.50603(12) & $3.927(12)$ & 7173\\  
   &    & 3.95 & 0.159452 & 0.00779 & 242 & $7.62(8)\cdot10^{-3}$ & 5.52831(11) & $3.720(12)$ & 8530\\  
   &    & 3.97 & 0.158926 & 0.00752 & 249 & $8.49(12)\cdot10^{-3}$& 5.55048(09) & $3.497(09)$ & 6518\\  
   &    & 4.01 & 0.157933 & 0.00718 & 265 & $1.05(1)\cdot10^{-2}$ & 5.59352(05) & $3.198(03)$ & 11240\\  
   &    & 4.05 & 0.157010 & 0.00680 & 281 & $1.25(1)\cdot10^{-2}$ & 5.63526(06) & $2.927(03)$ & 7264\\  
   &    & 4.10 & 0.155946 & 0.00639 & 302 & $1.51(1)\cdot10^{-2}$ & 5.68570(05) & $2.6556(14)$ & 6864\\  
   &    & 4.20 & 0.154073 & 0.00563 & 348 & $2.02(2)\cdot10^{-2}$ & 5.78199(06) & $2.2235(06)$ & 5231\\  
   &    & 4.35 & 0.151740 & 0.00466 & 428 & $2.87(1)\cdot10^{-2}$ & 5.91513(06) & $1.7331(02)$ & 5051\\  
\hline
 8 & 28 & 3.65 & 0.170250 & 0.01200 & 183 & $4.15(7)\cdot10^{-3}$ & -           & -           & 4100\\
   &    & 3.70 & 0.168062 & 0.01055 & 200 & $7.12(12)\cdot10^{-3}$& 5.1385(13)  & $7.589(25)$ & 4315\\  
   &    & 3.72 & 0.167220 & 0.01029 & 207 & $8.40(14)\cdot10^{-3}$& 5.22238(91) & $6.987(28)$ & 4943\\  
   &    & 3.74 & 0.166400 & 0.01004 & 215 & $9.89(15)\cdot10^{-3}$& 5.24896(71) & $6.426(18)$ & 5199\\  
   &    & 3.76 & 0.165607 & 0.00979 & 222 & $1.16(2)\cdot10^{-2}$ & 5.30285(45) & $5.910(27)$ & 4763\\  
   &    & 3.80 & 0.164111 & 0.00956 & 238 & $1.54(2)\cdot10^{-2}$ & 5.35463(22) & $5.239(12)$ & 4745\\
   &    & 3.85 & 0.162403 & 0.00893 & 258 & $2.00(2)\cdot10^{-2}$ & 5.41582(21) & $4.477(07)$ & 5130\\  
   &    & 3.90 & 0.160856 & 0.00821 & 280 & $2.44(2)\cdot10^{-2}$ & 5.47407(14) & $3.877(04)$ & 4199\\  
   &    & 3.93 & 0.159998 & 0.00801 & 293 & $2.72(2)\cdot10^{-2}$ & 5.50778(13) & $3.670(04)$ & 4468\\  
   &    & 3.95 & 0.159452 & 0.00779 & 302 & $2.92(2)\cdot10^{-2}$ & 5.53003(15) & $3.5080(19)$& 2640\\  
   &    & 3.97 & 0.158926 & 0.00752 & 311 & $3.07(3)\cdot10^{-2}$ & 5.55158(14) & $3.3380(20)$& 2688\\  
   &    & 4.01 & 0.157933 & 0.00718 & 330 & $3.43(2)\cdot10^{-2}$ & 5.59457(12) & $3.0964(13)$& 4200\\  
   &    & 4.10 & 0.155946 & 0.00639 & 377 & $4.23(3)\cdot10^{-2}$ & 5.68634(14) & $2.6127(08)$ & 2122\\  
   &    & 4.20 & 0.154073 & 0.00563 & 435 & $5.16(2)\cdot10^{-2}$ & 5.78232(12) & $2.1970(04)$ & 2689\\  
\hline
6  & 32 & 3.70 & 0.168062 & 0.01055 & 267 & $5.59(2)\cdot10^{-2}$ & 5.23893(18) & $5.8523(42)$& 6314\\  
   &    & 3.76 & 0.165607 & 0.00979 & 297 & $6.50(2)\cdot10^{-2}$ & 5.31365(14) & $5.0121(17)$& 5351\\  
   &    & 3.80 & 0.164111 & 0.00956 & 317 & $7.09(2)\cdot10^{-2}$ & 5.36179(13) & $4.6898(14)$& 4353\\  
   &    & 3.85 & 0.162403 & 0.00893 & 344 & $7.85(2)\cdot10^{-2}$ & 5.42080(11) & $4.1833(09)$& 4276\\  
   &    & 3.90 & 0.160856 & 0.00821 & 373 & $8.60(1)\cdot10^{-2}$ & 5.47772(10) & $3.6980(05)$& 6027\\  
   &    & 3.95 & 0.159452 & 0.00779 & 403 & $9.32(2)\cdot10^{-2}$ & 5.53282(11) & $3.3895(03)$& 4520\\  
   &    & 4.01 & 0.157933 & 0.00718 & 441 & $1.019(2)\cdot10^{-1}$& 5.59667(10) & $3.0112(03)$& 4054\\  
   &    & 4.05 & 0.157010 & 0.00680 & 468 & $1.078(2)\cdot10^{-1}$& 5.63790(10) & $2.7902(02)$& 3069\\  
   &    & 4.10 & 0.155946 & 0.00639 & 503 & $1.146(2)\cdot10^{-1}$& 5.68786(08) & $2.5554(02)$& 5168\\  
   &    & 4.20 & 0.154073 & 0.00563 & 580 & $1.285(2)\cdot10^{-1}$& 5.78372(10) & $2.1564(02)$& 3167\\
\hline
4  & 32 & 3.70 & 0.168062 & 0.01055 & 400 & $2.346(1)\cdot10^{-1}$& 5.27035(14) & $5.1616(06)$& 4353\\  
   &    & 3.76 & 0.165607 & 0.00979 & 445 & $2.442(1)\cdot10^{-1}$& 5.33934(14) & $4.5500(03)$& 4840\\  
   &    & 3.80 & 0.164111 & 0.00956 & 476 & $2.506(1)\cdot10^{-1}$& 5.38473(12) & $4.3062(02)$& 4714\\  
   &    & 3.85 & 0.162403 & 0.00893 & 517 & $2.587(1)\cdot10^{-1}$& 5.44025(12) & $3.8816(03)$& 3463\\ 
    \end{tabular}
 \end{center}
\caption{Simulation parameters for the C mass ensembles.}
\label{tab_Censemble}
\end{table*}
\begin{table*}
\footnotesize
 \begin{center}
    \begin{tabular}{c c|c|c|c|c|c|c|c|c}
       $N_\tau$ & $N_\sigma$ & $\beta$ &  $\kappa$  & $a\mu$ & $T [\mathrm{MeV}]$ & $\Real{(L)}$ &  $\vev{S_g}$& $\vev{S_f}~(\cdot10^{2})$ & TU  \\
       \hline\hline
10 & 24 & 3.76 & 0.165608 & 0.02218 & 178 & $8.5(4)\cdot10^{-4}$    & 5.29599(39) &  $13.561(09)$ &	9076\\
   &    & 3.80 & 0.164111 & 0.02100 & 190 & $1.21(4)\cdot10^{-3}$   & 5.34724(24) &  $12.185(08)$ & 	9391\\
   &    & 3.85 & 0.162401 & 0.01962 & 207 & $2.06(6)\cdot10^{-3}$   & 5.41016(18) &  $10.652(11)$ & 	9044\\
   &    & 3.90 & 0.160856 & 0.01833 & 224 & $3.64(10)\cdot10^{-3}$  & 5.47014(18) &  $9.3163(12)$ & 	9648\\
   &    & 3.93 & 0.159997 & 0.01759 & 234 & $4.98(15)\cdot10^{-3}$  & 5.50473(16) &  $8.6074(17)$ & 	7742\\ 
   &    & 3.97 & 0.158934 & 0.01666 & 249 & $7.25(13)\cdot10^{-3}$  & 5.54965(10) &  $7.7486(11)$ & 	11627\\
   &    & 3.99 & 0.158421 & 0.01621 & 257 & $8.47(16)\cdot10^{-3}$  & 5.57166(12) &  $7.3759(10)$ & 	8546\\   
   &    & 4.01 & 0.157933 & 0.01578 & 265 & $9.41(24)\cdot10^{-3}$  & 5.59281(14) &  $7.0487(21)$ & 	4059\\   
   &    & 4.05 & 0.157010 & 0.01524 & 281 & $1.22(2)\cdot10^{-2}$   & 5.63506(11) &  $6.5485(05)$ & 	7210\\
   &    & 4.10 & 0.155946 & 0.01431 & 302 & $1.45(3)\cdot10^{-2}$   & 5.68574(13) &  $5.9618(09)$ & 	3373\\    
   &    & 4.20 & 0.154073 & 0.01261 & 348 & $2.03(2)\cdot10^{-2}$   & 5.78192(08) &  $4.9791(02)$ & 	7059\\
   &    & 4.25 & 0.153241 & 0.01184 & 372 & $2.30(3)\cdot10^{-2}$   & 5.82768(12) &  $4.5758(02)$ & 	3377\\   
   &    & 4.35 & 0.151740 & 0.01043 & 428 & $2.87(3)\cdot10^{-2}$   & 5.91509(13) &  $3.8838(01)$ & 	3368\\   
\hline
8  & 20 & 3.65 & 0.170200 & 0.02517 & 183 & $3.03(4)\cdot10^{-3}$   & -           &  -            & 	21656\\
   &    & 3.70 & 0.168063 & 0.02406 & 200 & $4.51(7)\cdot10^{-3}$   & 5.21663(41) &  $15.762(14)$ & 	18240\\
   &    & 3.72 & 0.167219 & 0.02342 & 207 & $5.29(9)\cdot10^{-3}$   & 5.24389(37) &  $14.827(15)$ & 	16961\\
   &    & 3.74 & 0.166400 & 0.02279 & 215 & $6.46(8)\cdot10^{-3}$   & 5.27051(29) &  $13.942(14)$ & 	24771\\
   &    & 3.76 & 0.165608 & 0.02218 & 222 & $8.1(2)\cdot10^{-3}$    & 5.29755(30) &  $13.104(22)$ & 	18392\\
   &    & 3.78 & 0.164845 & 0.02158 & 230 & $9.7(2)\cdot10^{-3}$    & 5.32426(22) &  $12.272(18)$ & 	19091\\
   &    & 3.80 & 0.164111 & 0.02100 & 238 & $1.22(2)\cdot10^{-2}$   & 5.35106(20) &  $11.463(28)$ & 	33329\\
   &    & 3.82 & 0.163406 & 0.02044 & 246 & $1.44(2)\cdot10^{-2}$   & 5.37657(19) &  $10.777(17)$ & 	22399\\
   &    & 3.84 & 0.162730 & 0.01989 & 254 & $1.65(2)\cdot10^{-2}$   & 5.40132(18) &  $10.186(13)$ & 	20450\\
   &    & 3.85 & 0.162401 & 0.01962 & 258 & $1.80(2)\cdot10^{-2}$   & 5.41405(14) &  $ 9.861(10)$ & 	27895\\
   &    & 3.86 & 0.162080 & 0.01935 & 262 & $1.94(2)\cdot10^{-2}$   & 5.42623(13) &  $ 9.585(14)$ & 	24244\\
   &    & 3.87 & 0.161764 & 0.01909 & 267 & $2.02(2)\cdot10^{-2}$   & 5.43812(13) &  $ 9.329(14)$ & 	23330\\
   &    & 3.88 & 0.161456 & 0.01883 & 271 & $2.16(2)\cdot10^{-2}$   & 5.45008(15) &  $ 9.096(12)$ & 	21986\\
   &    & 3.90 & 0.160856 & 0.01833 & 280 &   $2.33(2)\cdot10^{-2}$ & 5.47311(10) &  $ 8.653(07)$ & 	23455\\
   &    & 3.92 & 0.160280 & 0.01784 & 288 &   $2.58(2)\cdot10^{-2}$ & 5.49627(10) &  $ 8.252(07)$ & 	22831\\
   &    & 3.94 & 0.159726 & 0.01736 & 297 &   $2.73(2)\cdot10^{-2}$ & 5.51844(14) &  $ 7.885(06)$ & 	 8812\\
   &    & 3.96 & 0.159193 & 0.01689 & 307 &   $2.93(2)\cdot10^{-2}$ & 5.54054(12) &  $ 7.548(05)$ & 	10974\\
   &    & 3.98 & 0.158681 & 0.01644 & 316 &   $3.12(2)\cdot10^{-2}$ & 5.56225(12) &  $ 7.234(04)$ & 	10889\\
   &    & 4.05 & 0.157010 & 0.01524 & 351 &   $3.80(2)\cdot10^{-2}$ & 5.63588(09) &  $ 6.4043(14)$& 	14180\\
   &    & 4.10 & 0.155952 & 0.01431 & 377 &   $4.24(2)\cdot10^{-2}$ & 5.68620(11) &  $ 5.8511(12)$& 	10233\\
   &    & 4.20 & 0.154073 & 0.01261 & 435 &   $5.17(2)\cdot10^{-2}$ & 5.78223(09) &  $ 4.9219(05)$& 	10861\\
   &    & 4.25 & 0.153238 & 0.01184 & 466 &   $5.62(2)\cdot10^{-2}$ & 5.82806(09) &  $ 4.5297(10)$& 	11573\\
   &    & 4.35 & 0.151740 & 0.01043 & 535 &   $6.52(2)\cdot10^{-2}$ & 5.91533(07) &  $ 3.8505(02)$& 	12089\\
\hline
6 & 16 & 3.65 & 0.170200 & 0.02517 & 243 & $4.29(5)\cdot10^{-2}$ & -           &  -  		  & 	10028\\
  &    & 3.70 & 0.168063 & 0.02406 & 267 & $5.48(6)\cdot10^{-2}$ & 5.23642(40) &  $13.341(14)$    & 	10198\\
  &    & 3.76 & 0.165608 & 0.02218 & 297 & $6.44(5)\cdot10^{-2}$ & 5.31246(25) &  $11.360(06)$    & 	13090\\
  &    & 3.80 & 0.164111 & 0.02100 & 317 & $7.03(4)\cdot10^{-2}$ & 5.36159(24) &  $10.301(05)$    & 	11773\\
  &    & 3.85 & 0.162401 & 0.01962 & 344 & $7.80(5)\cdot10^{-2}$ & 5.42026(22) & $ 9.187(03)$    & 	10516\\
  &    & 3.90 & 0.160856 & 0.01833 & 373 & $8.50(5)\cdot10^{-2}$ & 5.47736(19) & $ 8.2499(17)$   & 	11207\\
  &    & 3.97 & 0.158934 & 0.01666 & 415 & $9.53(5)\cdot10^{-2}$ & 5.55430(11) & $ 7.1530(06)$   & 	11109\\
  &    & 4.05 & 0.157010 & 0.01524 & 468 & $1.076(4)\cdot10^{-1}$& 5.63774(10) & $ 6.2516(06)$   & 	15926\\
\hline
6 & 20 & 4.20 & 0.154073 & 0.01261 & 580 & - 			 & 5.78357(11) &  $4.8317(04)$    & 	 7568\\
  &    & 4.35 & 0.151740 & 0.01043 & 713 & - 			 & 5.91643(14) &  $3.8220(03)$    & 	 3881\\
    \end{tabular}
 \end{center}
\caption{Simulation parameters for the D mass ensembles.}
\label{tab_Densemble}
\end{table*}
\clearpage
\end{appendix}

\bibliographystyle{apsrev}

\end{document}